\documentclass[a4paper, 11pt]{article}

\usepackage{graphicx}%
\usepackage{multirow}%
\usepackage{amsmath,amssymb,amsfonts}%
\usepackage{amsthm}%
\usepackage{mathrsfs}%
\usepackage[title]{appendix}%
\usepackage{xcolor}%
\usepackage{textcomp}%
\usepackage{manyfoot}%
\usepackage{booktabs}%
\usepackage{algorithm}%
\usepackage{algorithmicx}%
\usepackage{algpseudocode}%
\usepackage{listings}%
\usepackage{natbib}
\usepackage{rotating}
\usepackage{multicol}

\usepackage{bbm}

\newcommand{\Autoref}[1]{%
	\begingroup%
	\def\chapterautorefname{Chapter}%
	\def\sectionautorefname{Section}%
	\def\subsectionautorefname{Subsection}%
	\autoref{#1}%
	\endgroup%
}

\usepackage{caption}
\usepackage[a4paper,top=3cm,bottom=3.5cm,left=2.5cm,right=2.5cm,marginparwidth=1.75cm]{geometry}

\usepackage[colorlinks=true, allcolors=blue]{hyperref}

\title{Multivariate Spatio-temporal Modelling for Completing Cancer Registries and Forecasting Incidence}
\author{Garazi Retegui$^{1,2}$, Jaione Etxeberria$^{1,2}$, Mar{\'i}a Dolores Ugarte$^{1,2*}$\\
	\\
	\small $^1$ Department of Statistics, Computer Science and Mathematics, Public University of \\
	\small Navarre (UPNA), Arrosadia Campus, Pamplona, 31006, Navarra, Spain.\\
	\small $^2$ Institute for Advanced Materials and Mathematics (INAMAT2), Public University \\ 
	\small of Navarre (UPNA), Arrosadia Campus, Pamplona, 31006, Navarra, Spain.\\
	\small $^{*}$ Corresponding author.\\
	\\
	\small lola@unavarra.es\\ \\
	\date{}
}

\begin{document}
	\maketitle
	
	\begin{abstract}
		\noindent 	Cancer data, particularly cancer incidence and mortality, are fundamental to understand the cancer burden, to set targets for cancer control and to evaluate the evolution of the implementation of a cancer control policy. However, the complexity of data collection, classification, validation and processing result in cancer incidence figures often lagging two to three years behind the calendar year. In response, national or regional population-based cancer registries (PBCRs) are increasingly interested in methods for forecasting cancer incidence.
		However, in many countries there is an additional difficulty in projecting cancer incidence as regional registries are usually not established in the same year and therefore cancer incidence data series between different regions of a country are not harmonised over time.
		This study addresses the challenge of forecasting cancer incidence with incomplete data at both regional and national levels.
		To achieve this, we propose the use of multivariate spatio-temporal shared component models that jointly model mortality data and available cancer incidence data. We evaluate the performance of these multivariate models using lung cancer incidence data and the corresponding number of deaths reported in England for the period 2001-2019.  Model performance was assessed using different predictive measures to select the best model.
	\end{abstract}
	\textbf{Keywords: } Cancer registries, Disease mapping, INLA, Predictions, Shared component models  

 \defcitealias{world2002}{WHO, 2002}
\section{Introduction} \label{Section:Introduction}
Cancer control aims to reduce the incidence, morbidity and mortality of cancer, and to advocate for cancer patients through the systematic implementation of evidence-based interventions in prevention, early detection, treatment and palliative care \citepalias{world2002}. Therefore, cancer data are essential to understand the current situation, to set targets for cancer control and to evaluate the evolution of the implementation of a cancer control policy. In this context, national or regional population-based cancer registries (PBCRs) are responsible for assessing the current magnitude of the cancer burden and its likely future evolution \citep{parkin2008}. Various statistics are available to assess the burden of cancer, such as incidence, to describe the number of new cases that will require some kind of medical treatment, and mortality, to evaluate the effectiveness of screening, early diagnosis and treatment programmes. However, collecting cancer incidence data presents significant challenges. The complexity arises from multiple factors, including the need to gather data from diverse sources, the change of disease classification and coding systems over time, the dynamic nature of criteria for defining new cancer cases, and the difficulty in differentiating between new cases and recurrences or metastases of existing cases \citep{parkin2006}.
Therefore, the complexity of data collection, classification, validation and processing means that cancer incidence figures are available two to three years behind the calendar year. In this context, PBCRs are highly interested in using methods to forecast cancer incidence figures and a variety of methods have been proposed in the literature to produce such forecasts. For example, simple linear \citep{hakulinen1994} and non-linear models \citep{dyba1997}, join-point regression models \citep{kim2000}, generalized additive models \citep{clements2005} or Lee-carter (LC) models \citep{lee1992}. But the methods commonly used in the literature to get projections are based on Age-Period-Cohort (APC) models \citep{holford1983}. For instance, Nordic countries employ the approach known as \textit{Nordpred} proposed by \cite{engeland1993} to forecast national cancer incidence \citep{moller2002}. These APC methods are applied to national incidence projections in countries where the PBCRs cover 100\% of the population or in areas with a large historical collection of cancer incidence data. However, in many large countries, an additional challenge arises when forecasting cancer incidence. Regional cancer registries, responsible for collecting and identifying cancer cases in specific areas, are often established at different times. This leads to incomplete and non-harmonized data series across regions, especially at the beginning of data collection. Consequently, it becomes difficult to use aggregated national data and apply the previously mentioned approaches.
In this case, the estimation and projection of national cancer incidence are generally performed in two steps   \citep[see for example][]{galceran2017}. First, national incidence estimates for each year in the data series are derived using the Incidence/Mortality ratio \citep{sung2021}. Then, projections are obtained using the above mentioned techniques.  \cite{retegui2023} propose an alternative approach to estimate cancer incidence in areas without registries. They use multivariate spatial models to combine cancer mortality data, readily available from statistical offices or cancer registries, with existing incidence data. This integration improves the accuracy of cancer incidence forecasts. These models 
can be extended to a spatio-temporal context, allowing for area-level projections that accommodate non-harmonized data across regions, as well as projections at the national level.

In this paper, we examine the effectiveness of multivariate spatio-temporal models in addressing the double challenge of forecasting short-term cancer incidence and completing incomplete cancer registry data series, with the ultimate goal of generating national cancer incidence projections. We investigate both shared component models, SCMs, \citep[][]{held2005} and M-models \citep{botella2015}. However, in line with \cite{retegui2023}, our preliminary analyses show that SCMs consistently outperformed M-models in the real-world data considered here.  Given this superior performance, this paper focuses on SCMs. SCMs provide a straightforward approach to modelling multiple health outcomes when a priori dependence, such as that between cancer incidence and mortality, is assumed. To address the challenges posed by incomplete data series, we explore various SCM specifications, incorporating spatial, temporal, and spatio-temporal shared components.
We evaluate the performance of these multivariate models using lung cancer data for England from 2001 to 2019, alongside corresponding mortality data, and employ a hold-out validation strategy to assess models' performance. This involves removing data from specific time periods and regions, and then evaluating the models's ability to predict these values against the actual registered data.

The rest of the paper is laid out as follows.   
\Autoref{Section:EDA} presents an exploratory data analysis of lung cancer incidence and mortality data in England from 2001 to 2019. \Autoref{Section:MultivariateModels} describes the proposed spatio-temporal multivariate models with spatial, temporal, or spatio-temporal shared components for short-term forecasting. Details on implementing these models using Integrated Nested Laplace Approximations (INLA) are also provided.
\Autoref{Section:Validation} outlines the cross-validation procedure employed to  assess the models' predictive performance and presents results using real data at regional level.  \Autoref{Section:National} extends these findings by reporting the corresponding national-level results. The paper concludes with a discussion.

\section{Exploratory data analysis} \label{Section:EDA}

\begin{figure}[!b]
	\begin{center}
		\includegraphics [width=7cm]{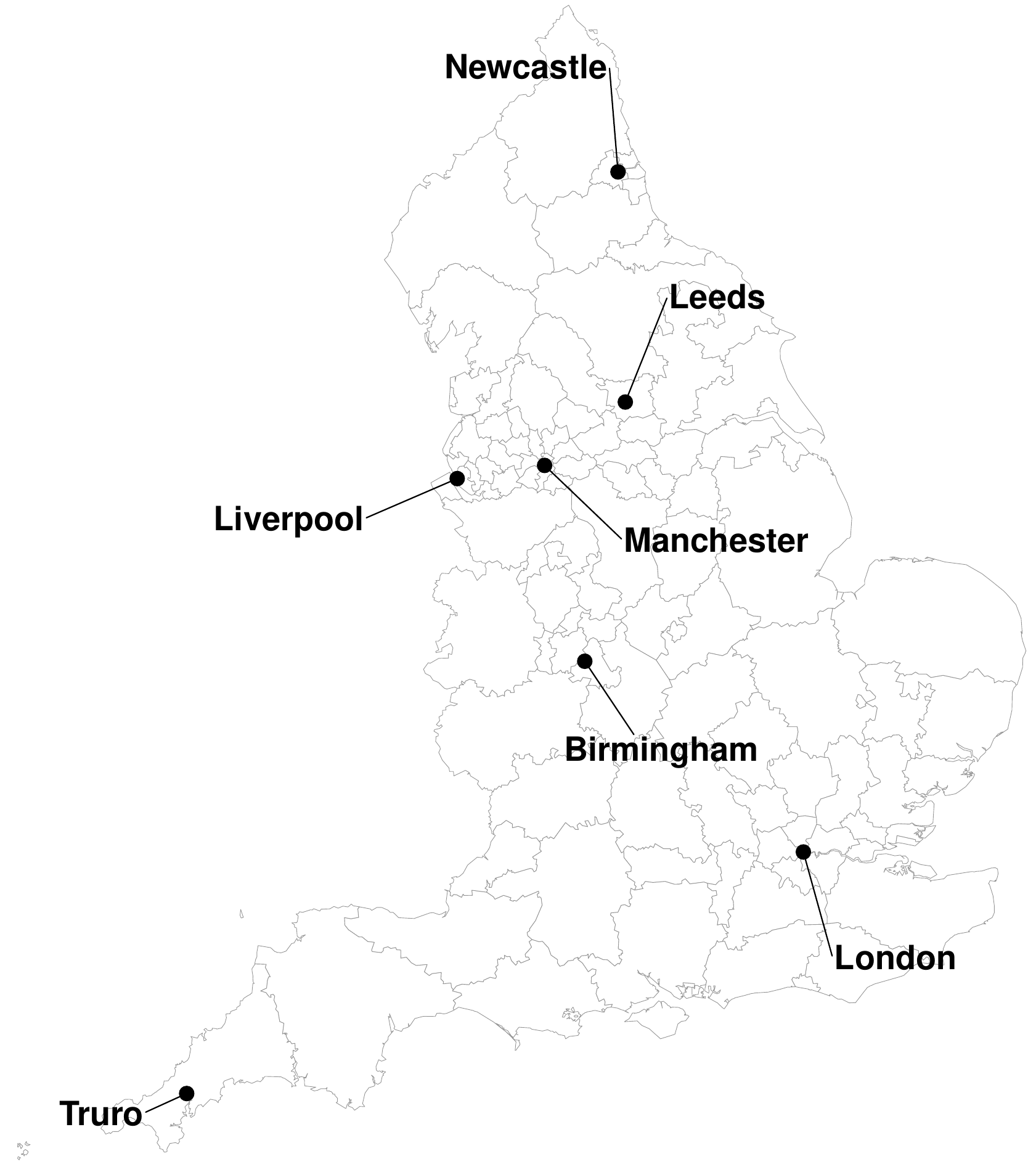}
	\end{center}
	\caption{106 English clinical commissioning groups. \label{fig4.1}}
\end{figure}

\begin{figure}[!t]
	\begin{center}
		\includegraphics [width=11cm]{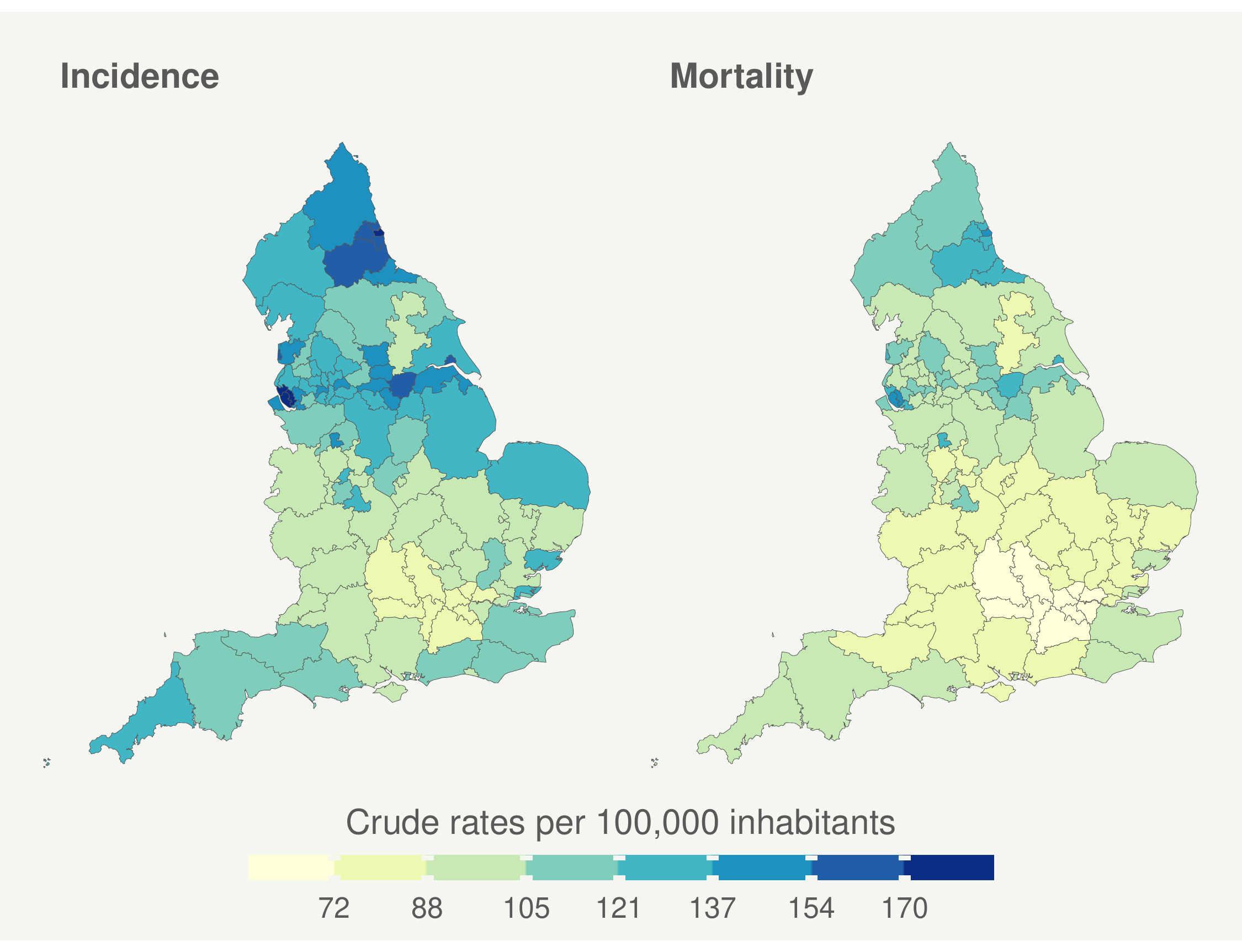}
	\end{center}
	\caption{Spatial distribution of crude incidence and mortality rates per 100,000 inhabitants for lung cancer during the period 2001-2019. \label{fig4.2}}
\end{figure}

This study uses lung cancer incidence and mortality data for males in England. The data is stratified by the 106 clinical commissioning groups (CCG) (see \autoref{fig4.1}) and covers the period from 2001 to 2019. During the 19 years of the study period, a total of 371,835 incident cases and 298,929 deaths were recorded. \autoref{fig4.2} shows the geographical patterns of crude incidence and mortality rates per 100,000 inhabitants. The spatial patterns of incidence and mortality are very similar. Crude rates range from 74 to 185 incident cases, and from 57 to 149 deaths per 100,000 inhabitants. In general, the highest crude rates are predominantly observed in the northern regions and areas near Liverpool and south of Leeds. Conversely, the lowest rates for both incidence and mortality are observed in regions south and west of London.

\begin{figure}[!b]
	\begin{center}
		\includegraphics [width=11cm]{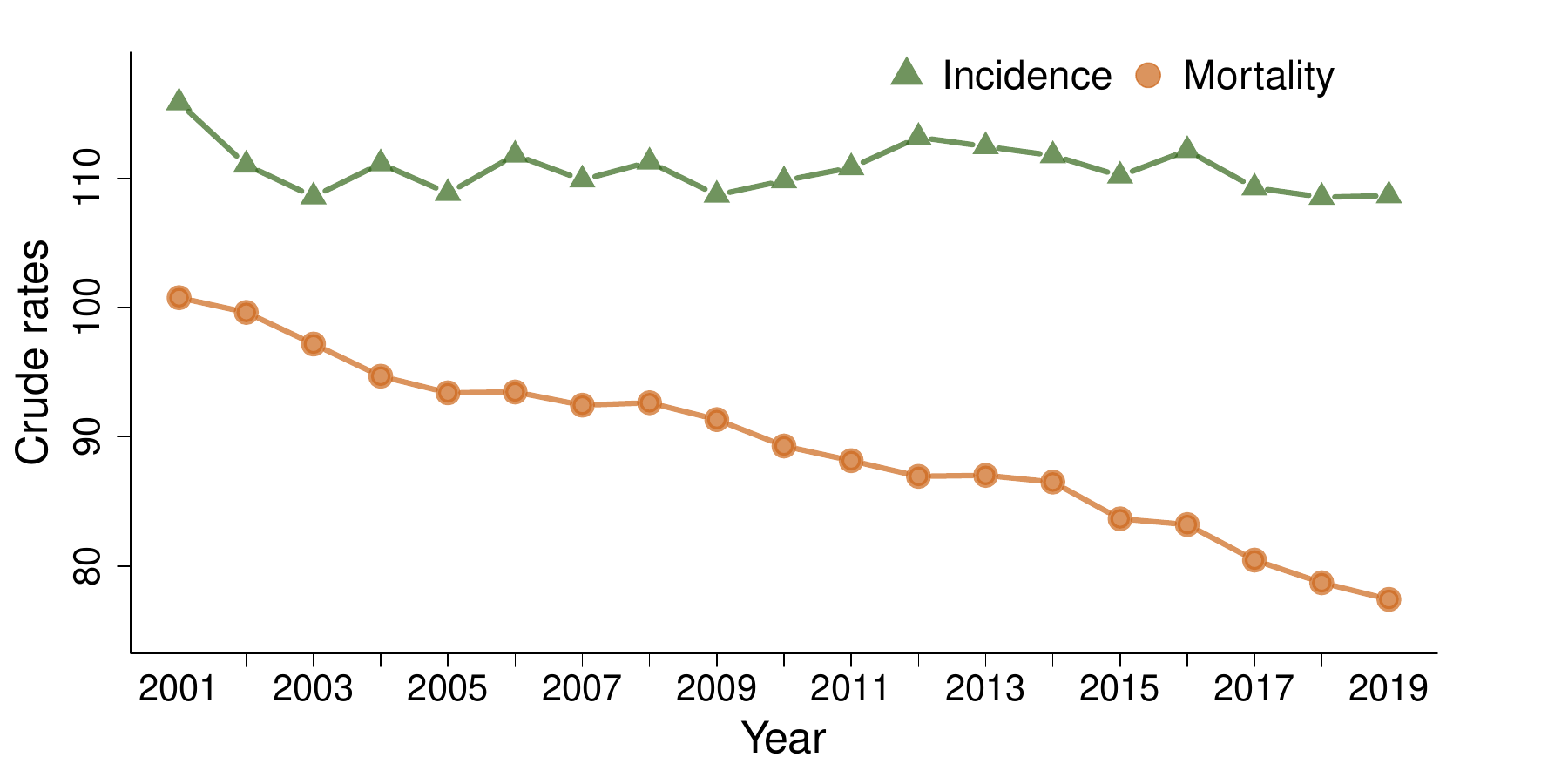}
	\end{center}
	\caption{Temporal trend of crude incidence and mortality lung cancer rates per 100,000 inhabitants throughout the study period (2001-2019).\label{fig4.3}}
\end{figure}

\autoref{fig4.3} displays the temporal trends of crude incidence and mortality rates per 100,000 inhabitants during the study years. We observe a relatively stable incidence rate throughout the study period, ranging between 105 and 120. In contrast, the mortality rate decreases linearly, starting around 100 crude rate and decreasing until 70. Thus, distinct temporal trends were observed between lung cancer incidence and mortality.

%

\begin{figure}[!t]
	\begin{center}
		\scalebox{0.3}{\includegraphics[page=1]{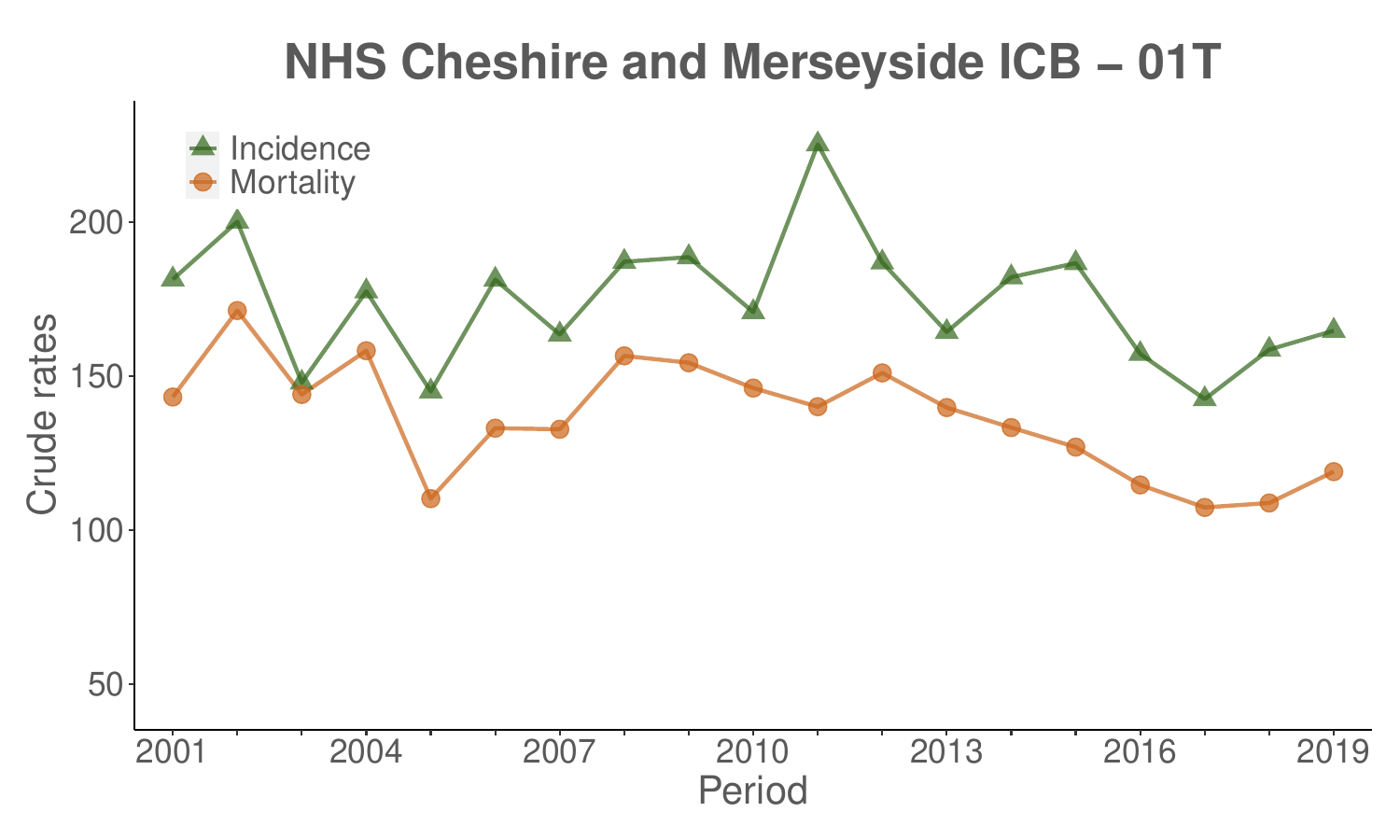}}\hspace{0.5cm}
		\scalebox{0.3}{\includegraphics[page=2]{Fig/Figure4.pdf}}\\ \vspace*{0.7cm}
		
		\scalebox{0.3}{\includegraphics[page=3]{Fig/Figure4.pdf}}\hspace{0.5cm}
		\scalebox{0.3}{\includegraphics[page=4]{Fig/Figure4.pdf}}
	\end{center}
	\caption{\label{fig4.4} Evolution of crude area-specific incidence and mortality lung cancer rates per 100,000 inhabitants in NHS Cheshire and Merseyside ICB - 01T, NHS South Yorkshire ICB - 02X, NHS Cornwall and The Isles Of Scilly ICB - 11N and NHS North Central London ICB - 93C.}
\end{figure}

\autoref{fig4.4} illustrates the temporal trends in incidence and mortality rates for four specific CCG in England, namely NHS Cheshire and Merseyside ICB - 01T, NHS South Yorkshire ICB - 02X, NHS Cornwall and The Isles Of Scilly ICB - 11N and NHS North Central London ICB - 93C. These CCG correspond to areas encompassing Liverpool, Doncaster (located south-east of Leeds), Truro and London, respectively. The incidence and mortality rates generally exhibit similar temporal trends for each area. Notably, variations are observed both in the ratio between incidence and mortality rates within specific years and in trends during certain periods for NHS Cheshire and Merseyside ICB. For instance, variation in the ratio is evident in the year 2011 and distinct trends are observed between the periods 2013 and 2015. Moreover, in general, the level of association between incidence and mortality remains remarkably consistent over time for each area.

\section{Multivariate spatio-temporal models} \label{Section:MultivariateModels}
In this section, we define different spatio-temporal SCMs to predict cancer incidence using mortality data.
Let $O_{itd}$ and $n_{itd}$ be the number of observed cases and population at risk, respectively, for the $i$th geographical area ($i=1,2,\dots,A$), the $t$th year ($t=1,\dots,T$) and the $d$th health outcome ($d=I$ (incidence), $d=M$ (mortality)).
Conditional on the area, time and health outcome-specific rate, denoted by $r_{itd}$, we assume that the number of observed cases follows a Poisson distribution with mean $\mu_{itd}=n_{itd}r_{itd}$. Namely	
\begin{eqnarray}\label{eq1}
	O_{itI}|r_{itI}&\sim& Poisson\left(\mu_{itI}=n_{itI}r_{itI}\right); \quad \quad \quad \log \mu_{itI}=\log n_{itI}+\log r_{itI}, \nonumber \\
	O_{itM}|r_{itM}&\sim& Poisson\left(\mu_{itM}=n_{itM}r_{itM}\right); \quad  \log \mu_{itM}=\log n_{itM}+\log r_{itM}.
\end{eqnarray}

\noindent Then, we model  $\log r_{iI}$ and $\log r_{iM}$ using different proposals. For ease of reading, we describe here only the best-performing model for each of the three types of models considered: models with only shared spatial components (Type 1); models with both shared spatial and temporal components (Type 2), and models with shared spatial and spatio-temporal components (Type 3). As a baseline, we also evaluated a model assuming additive spatial and temporal effects without spatio-temporal interaction. However, this model exhibited significantly poorer performance across the evaluation metrics and was therefore not included in the main analysis.

Let us first start with Model 1 (the best-performing model within the Type 1 category) including a health outcome-specific intercept $\alpha_d$, a shared component spatial term $\kappa_{i}$, a spatially unstructured random effect for mortality $u_{i}$, a time effect specific for each health outcome  $\gamma_{td}$ and spatio-temporal interactions specific for incidence and mortality $\chi_{itd}$. Namely
\begin{eqnarray*}
	&\textrm{Model 1:}&\log r_{itI} =\alpha_I + \delta \kappa_{i} + \gamma_{tI} + \chi_{itI}, \nonumber \\
	&&\log r_{itM} =\alpha_M + \frac{1}{\delta} \kappa_{i} + u_{i} + \gamma_{tM} + \chi_{itM},
\end{eqnarray*}
where $\delta$ is a scaling parameter. The spatio-temporal interaction will have the same structure for both health outcomes but the amount of smoothing for each health outcome can be the same or different.

The following prior distributions are assumed for the different terms of Model 1:
\begin{eqnarray}
	&&p(\boldsymbol{\kappa}) \propto \exp\left(\frac{-\tau_{\kappa}}{2} \boldsymbol{\kappa}^{'} \mathbf{R}_{\kappa} \boldsymbol{\kappa} \right), \nonumber \\
	&&\delta \sim \textrm{Gamma}(10,10), \nonumber \\
	&&p(\mathbf{u}) \propto \exp\left(\frac{-\tau_u}{2}\boldsymbol{u}^{'} \mathbf{I}_A \boldsymbol{u}\right), \nonumber\\
	&&p(\mathbf{\boldsymbol{\gamma}_d}) \propto \exp\left(\frac{-\tau_{\gamma_d}}{2} \boldsymbol{\gamma}_d^{'} \mathbf{R}_{\gamma} \boldsymbol{\gamma}_d \right), \nonumber\\
	&&p(\boldsymbol{\chi}_{d}) \propto \exp\left(\frac{-\tau_{\chi_d}}{2} \boldsymbol{\chi}_{d}^{'} \mathbf{Q}_{\chi} \boldsymbol{\chi}_{d} \right),\nonumber
\end{eqnarray}

\begin{table}[b!]
	\centering
	\caption{Specification of the four types of space-time interaction proposed by \cite{knorr2000}, along with the identifiability constraints defined for each interaction type \citep{goicoa2018}. \label{tab3.0}}
	\begin{tabular}{lcccccc}
		\toprule
		Space-time interaction &\multicolumn{2}{c}{Structure in} && $\mathbf{Q}_{\chi}$ && Identifiability constraints \\
		&Time & Space &&&&\\
		\midrule
		Type I &  &  && $\mathbf{I}_{T} \otimes \mathbf{I}_{A}$ && $\sum_{i=1}^{A} \sum_{t=1}^{T} \chi_{it} = 0$  \\
		\multicolumn{7}{c}{ }\\
		Type II & \checkmark &  &&  $\mathbf{R}_{\gamma} \otimes \mathbf{I}_A$ && $ \sum_{t=1}^{T} \chi_{it} = 0 , \forall i$\\
		\multicolumn{7}{c}{ }\\
		Type III &  & \checkmark&& $\mathbf{I}_T \otimes \mathbf{R}_{\kappa}$ && $\sum_{i=1}^{A} \chi_{it} = 0, \forall t$\\
		\multicolumn{7}{c}{ }\\
		\multirow{2}{*}{Type IV}&\multirow{2}{*}{\checkmark}& \multirow{2}{*}{\checkmark}&&\multirow{2}{*}{$ \mathbf{R}_{\gamma} \otimes \mathbf{R}_{\kappa}$ } && $\sum_{i=1}^{A} \chi_{it} = 0, \forall t$ \\
		&&&&&& $\sum_{t=1}^{T} \chi_{it} = 0, \forall i$\\
		\bottomrule
	\end{tabular}
\end{table}

\noindent where $\mathbf{R}_{\kappa}$ is the spatial neighborhood matrix defined by \cite{besag1991} considering that two areas are neighbors if they share a common border, $\mathbf{I}_{A}$ is the identity matrix of size $A\times A$, $\mathbf{R}_{\gamma}$ is determined by the temporal structure matrix of a first order random walk \citep[see][p. 95]{rue2005} and $\mathbf{Q}_{\chi}$ is the structure matrix that represents any of the four spatio-temporal interaction types proposed by \cite{knorr2000}.
The structure matrices $\mathbf{Q}_{\chi}$ for the different interaction types are summarized in \autoref{tab3.0} together with the identifiability constraints needed to implement them.

Our motivating context involves incomplete spatio-temporal data. This may present challenges when estimating the temporal incidence effect ($\boldsymbol{\gamma_{tI}}$) in Model 1. Then, we consider Model 2 (the best performing model of Type 2 models) adding a shared term for the temporal effect  \citep{etxeberria2023}. Namely
\begin{eqnarray*}
	&\textrm{Model 2:}&\log r_{itI} =\alpha_I + \delta \kappa_{i} + \varsigma \gamma_{t} +\chi_{itI}, \nonumber \\
	&&\log r_{itM} =\alpha_M + \frac{1}{\delta} \kappa_{i} + u_{i} + \frac{1}{\varsigma} \gamma_t  +\chi_{itM},
\end{eqnarray*}
where $\varsigma$ represents an additional scaling parameter, and $\gamma_t$ denotes the shared temporal effect. We consider the following prior distributions for these parameters.
\begin{eqnarray*}
	\varsigma&\sim& \textrm{Gamma}(10,10),  \\
	p(\mathbf{\boldsymbol{\gamma}}) &\propto& \exp\left(\frac{-\tau_{\gamma}}{2} \boldsymbol{\gamma}^{'} \mathbf{R}_{\gamma} \boldsymbol{\gamma} \right).
\end{eqnarray*}

While incidence and mortality rates generally exhibit similar temporal trends within each area, our exploratory data analysis revealed distinct temporal trends when comparing incidence and mortality globally. This observation led us to focus on the best-performing model from the Type 3 category (Model~3), which incorporates shared spatio-temporal interactions with time-varying scaling parameters. 
This model was introduced by \cite{retegui2024} to analyze the spatio-temporal patterns of rare cancers but its predictive performance has not been evaluated yet. Model~3 seeks to enhance prediction accuracy by sharing spatio-temporal information between cancer incidence and mortality data. Namely
\begin{eqnarray*}
	&\textrm{Model 3:}&\log r_{itI} =\alpha_I + \delta \kappa_{i} + \varrho_t \chi_{it}, \nonumber \\
	&&\log r_{itM} =\alpha_M + \frac{1}{\delta} \kappa_{i} + u_i + \gamma_{Mt} + \frac{1}{\varrho_t} \chi_{it}
\end{eqnarray*}
where $\varrho_t$ is a scaling parameter for time $t$ and $\boldsymbol{\chi}$ is the shared spatio-temporal interaction. Note that Model~3 does not include a temporal incidence effect, $\boldsymbol{\gamma_{tI}}$,  as the incomplete data makes its estimation challenging. Nevertheless, the shared spatio-temporal interaction implicitly captures the underlying temporal trend. We consider the following prior distributions for these parameters
\begin{eqnarray*}
	\varrho_t &\sim& \textrm{Gamma}(10,10)\quad \quad t = 1, \dots, l   \textrm{ where }  1\leq l\leq T,  \\ p(\boldsymbol{\chi}) &\propto& \exp\left(\frac{-\tau_{\chi}}{2} \boldsymbol{\chi}^{'} \mathbf{Q}_{\chi} \boldsymbol{\chi} \right),
\end{eqnarray*}
\noindent where $\mathbf{Q}_{\chi}$ has any of the structure matrices defined by \cite{knorr2000}.

\subsection{Model fitting and inference} \label{Section:MultivariateModels.ModelFitting}
We use the integrated nested Laplace approximation (INLA) technique \citep{rue2009} to fit the models described above. INLA can be implemented in the free software R through the R-package R-INLA \citep[][]{martino2009, martino2014}.  We assume uniform vague prior distributions for the standard deviations $\sigma_{\kappa} = 1/\sqrt{\tau_{\kappa}}, \sigma_u  = 1/\sqrt{\tau_{u}}$, $\sigma_{w} = 1/\sqrt{\tau_{w}}$, $\sigma_{\gamma}  = 1/\sqrt{\tau_{\gamma}}$, $\sigma_{\Phi} = 1/\sqrt{\tau_{\Phi}}$ and $\sigma_{\psi} = 1/\sqrt{\tau_{\psi}}$.
As the multivariate spatio-temporal models suffer from identifiability issues, we impose sum-to-zero constraints on the spatial, temporal and spatio-temporal random effects.
For more details about the required constraints for each case (see \autoref{tab3.0}).

In this work, one of our focus lies in forecasting cancer incidence rates. Forecasting involves fitting a model with missing data, specifically for the observations corresponding to the years we aim to forecast. Consequently, it is necessary to assign \texttt{NA} values to the observations that will be forecasted. \texttt{R-INLA} enables handling missing observations in the response variable and computes the posterior marginal for the corresponding linear predictor, $\log r_{itd}$.
If we are interested in the posterior predictive density of the missing observation, post-processing of the result is necessary \citep[see for example][]{martino2019}.
Additionally, it is important to note that besides forecasting, the incidence counts $O_{itI}$ are not available for certain geographical areas in specific years.  Consequently, it is necessary to assign \texttt{NA} values to the observations that are missing. 
Computations were performed using simplified Laplace approximation strategy in the stable version of R-INLA INLA\_22.12.16. The full code and data to reproduce results are available at \url{https://github.com/spatialstatisticsupna/Forecasting_Cancer_Incidence}.

\section{Model validation} \label{Section:Validation}

In this section, we evaluate the predictive performance of the models introduced in \autoref{Section:MultivariateModels}.
We use lung cancer incidence and mortality data for males in England, as presented in \autoref{Section:EDA}. The models' evaluation involves excluding data from specific areas and years, and then comparing the models' predictions with the recorded data.

\subsection{Model validation procedure}

\begin{figure}[!b]
	\begin{center}
		\includegraphics [width=13.3cm]{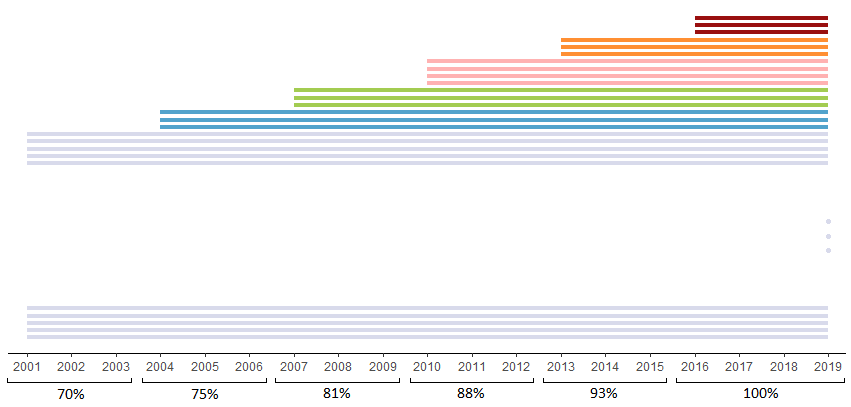} \\ \vspace*{1cm}
		
		\includegraphics [width=4cm,  page=1]{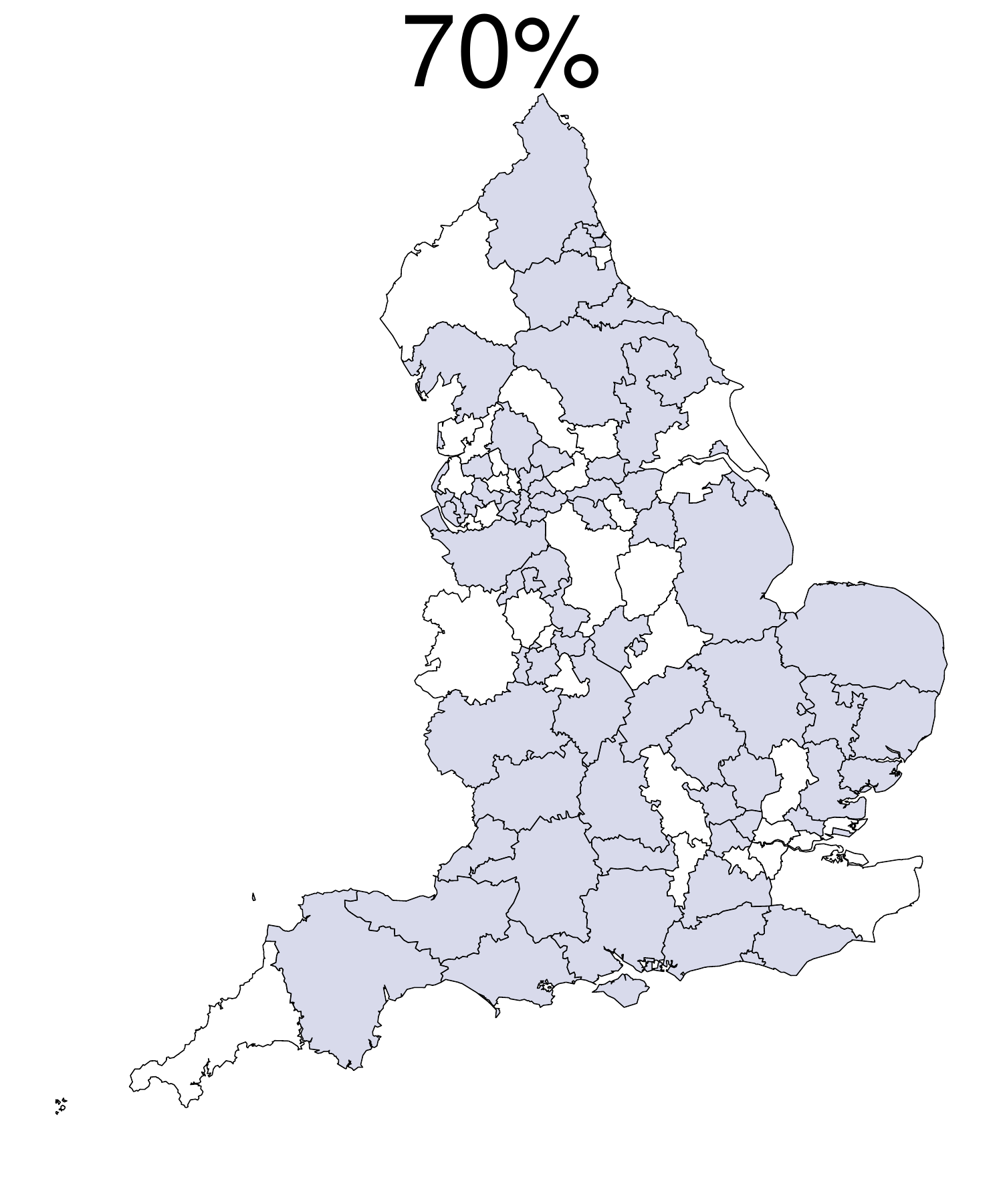} \hspace{0.5cm}
		\includegraphics [width=4cm,  page=2]{Fig/Figure5_2.pdf} \hspace{0.5cm}
		\includegraphics [width=4cm,  page=3]{Fig/Figure5_2.pdf}\\ \vspace{0.5cm}
		
		\includegraphics [width=4cm,  page=4]{Fig/Figure5_2.pdf} \hspace{0.5cm}
		\includegraphics [width=4cm,  page=5]{Fig/Figure5_2.pdf}  \hspace{0.5cm}
		\includegraphics [width=4cm,  page=6]{Fig/Figure5_2.pdf}
	\end{center}
	\caption{Graph showing the percentage of  areas with data availability for each year (top) and maps of England highlighting regions with available incidence data (colored areas) for various population coverage percentages. Lilac colouring indicates areas with complete data (no missing years), while blue, green, pink, orange, and garnet represent areas with 3, 6, 9, 12, and 15 years of missing data, respectively. \label{fig4.5}}
\end{figure}

To evaluate the predictive performance of multivariate models, we withhold incidence data from specific regions and time periods, while retaining complete mortality data.  The strategy employed to simulate missing incidence data is informed by real-world scenarios, such as those documented in Germany \citep{Robert2020} and Switzerland \citep{NACR}.
The data series exhibit the lowest percentage of areas with complete data at the beginning, gradually increasing until reaching 100\% coverage of incidence data. To simulate this situation, we consider that at the beginning of the time series, incidence data is available for only 70\% of the areas. As the time series progresses, we gradually incorporate new areas as known. Specifically, the percentage of available areas increases to 75\%, 81\%, 88\%, 93\% and 100\%. Moreover, these percentages of available incidence data change every three years. For the years 2001-2003, 70\% of the incidence data is available, for the years 2004-2006, 75\%, and following this pattern, from 2016 onwards, 100\% of the incidence data is available. \autoref{fig4.5} (top panel) shows the percentage of available incidence data at each time point. The areas from which incidence data are removed are randomly selected. For more details on the removed areas, please see \autoref{fig4.5} (bottom panel).
After removing incidence data from specific CCGs, as outlined above, multivariate models (\Autoref{Section:MultivariateModels}) are employed to jointly model mortality data and the remaining incidence data. To evaluate model performance, a cross-validation approach is implemented. More specifically, three-year-ahead predictions of cancer incidence were generated as follows: predictions for 2014 were based on data from 2001 to 2011, predictions for 2015 were based on data from 2001 to 2012, and so forth \citep[similar to the methodology employed by][]{ghosh2007}. A similar approach was adopted for one- and two-year-ahead predictions. \autoref{fig4.6} in the \autoref{Appendix:Results} illustrates the procedure. A summary of the hyperparameters across the six validation periods is provided in \autoref{AppendixA}.

In the final stage of our analysis, we assess the performance of the various multivariate models using three distinct metrics: the absolute relative bias (ARB),  the Dawid-Sebastiani score \citep[DSS, ][]{dawid1999,gneiting2007} and the interval score \citep[IS, ][]{gneiting2007}.  These metrics, along with their corresponding means, were computed for a specific subset of areas (W) and years (V). Namely

\begin{eqnarray*}
	ARB_{it}&=&\frac{|\hat{r}_{itI}-r_{itI}|}{r_{itI}}, \quad i \in W, \quad t \in V,\\
	DSS_{it}&=&
	\left(\frac{C_{itI}-E(\hat{C}_{itI})}{sd(\hat{C}_{itI})}\right)^2 + 2\log\left(sd(\hat{C}_{itI})\right), \quad i\in W, \quad t \in V,\\
	IS^{it}_\beta \left(l,u;r_{itI}\right)&=& (u-l)+\frac{2}{\beta}(l-r_{itI})\mathbbm{1}\left\{r_{itI}<l\right\} + \frac{2}{\beta}(r_{itI}-u)\mathbbm{1}\left\{r_{itI}>u\right\}, \quad i \in W, \quad t \in V, \nonumber
\end{eqnarray*}

\begin{eqnarray*}
	MARB&=&\frac{1}{|V|}\frac{1}{|W|}\sum_{t\in V}\sum_{i \in W}ARB_{it},\\
	DSS&=&\frac{1}{|V|}\frac{1}{|W|}\sum_{t\in V}\sum_{i\in W}DSS_{it},\\
	IS&=&\frac{1}{|V|}\frac{1}{|W|}\sum_{t\in V}\sum_{i \in W}IS^{it}_\beta \left(l,u;r_{iI}\right),
\end{eqnarray*}
where $i$ is the area, $t$ is the year, $\hat{r}_{itI}$ is the posterior mean of the fitted incidence rate for area $i$ and year $t$, $r_{itI}$ is the observed rate,  $\hat{C}_{itI}$ is the predictive incidence count for area $i$ and year $t$ and $C_{itI}$ is the observed count. The mean and standard deviation of $\hat{C}_{itI}$ are computed from the posterior distribution of $\hat{C}_{itI}$ \cite[see][]{riebler2017}. Additionally, $l=q_{itI;\beta/2}$ and $u=q_{itI;1-\beta/2}$ are the $\beta/2$ and $1-\beta/2$ quantiles of the posterior distribution of the fitted incidence rate for area $i$ and time $t$, and $\mathbbm{1}\left\{.\right\}$ is the indicator function that takes value 1 if the event in brackets is true and 0 otherwise. Here, $|V|$ represents the number of years in V and $|W|$ the number of areas in W.

\subsection{Results} \label{Section:Results}
The analyses across all interaction types and various scale parameter values reveal that Type~II interaction consistently produces the best results for all models. Notably, for Model~3, a single temporal scale parameter yields the best performance (results not shown here to save space).
We categorize the results into two headings: predicting area-specific incomplete time series and short-term forecasting.

\subsubsection{Predicting area-specific incomplete time series}

\begin{table}[!b]
	\begin{center}
		\caption{\label{tab4.1}
			MARB, DSS and IS values for each model across three-year periods with varying levels of incidence data availability:2001-2003 (70\% available), 2004-2006 (75\% available), 2007-2009 (81\% available), 2010-1012 (88\% available) and 2013-2015 (93\% available).}
		\resizebox{\textwidth}{!}{
			\begin{tabular}{c|rrrcrrrcrrr}
				\toprule
				\multicolumn{1}{c}{ }&\multicolumn{3}{c}{\bf Model 1} & &\multicolumn{3}{c}{ \bf Model 2} &&\multicolumn{3}{c}{\bf Model 3}\\
				\cline{2-12}
				&&&&&&&&&&&\\[-0.8em]
				& MARB&  DSS&  IS&& MARB&  DSS&  IS&& MARB&  DSS&  IS\\
				\midrule
				&&&&&&&&&&&\\[-1em]
				2001-2003 (70\%) & 0.103  & 7.160 & 139.600  && 0.105 & 6.867 & 141.170  && 0.088 & 6.482 & 112.908\\
				&&&&&&&&&&&\\[-1em]
				2004-2006 (75\%) & 0.104 & 6.754 & 177.818 && 0.108 & 6.541 & 177.042  && 0.097 & 6.267  & 166.185\\
				&&&&&&&&&&&\\[-1em]
				2007-2009 (81\%) & 0.094 & 6.764 & 177.294 && 0.098 & 6.609 & 175.505 && 0.095 & 6.396 & 162.780\\
				&&&&&&&&&&&\\[-1em]
				2010-2012 (88\%) & 0.072 & 6.690 & 92.434  && 0.074 & 6.484 & 91.224 && 0.073 & 6.244 & 95.814\\
				&&&&&&&&&&&\\[-1em]
				2013-2015 (93\%) & 0.089 & 6.056 & 145.905 && 0.092 & 6.055 & 157.767 && 0.087 & 5.684 & 141.298\\
				\bottomrule
			\end{tabular}
			}
	\end{center}
\end{table}

\begin{figure}[!b]
	\begin{center}
		\includegraphics [width=14.75cm, page=1]{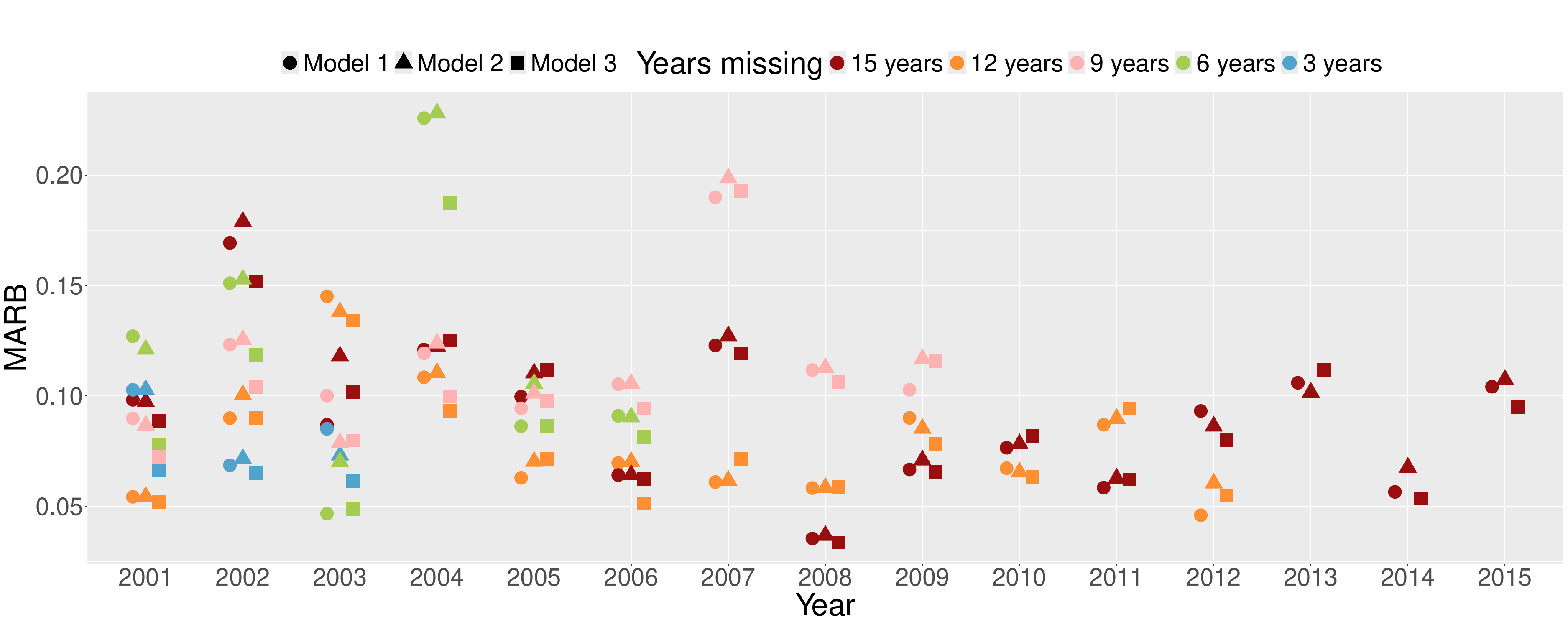} \\ \vspace*{0.5cm}
		
		\includegraphics [width=14.75cm, page=2]{Fig/Figure7.pdf}
	\end{center}
	\caption{Average values across validation periods for ARB and DSS, represented for each year of the time series and each group of areas based on the duration of missing data. Different shapes indicate the models used: dots for Model~1, triangles for Model~2, and squares for Model~3. Colours represent different groups of areas based on missing data duration: blue for areas with 3 years missing, green for 6 years, pink for 9 years, orange for 12 years, and garnet for 15 years. \label{fig4.9} }
\end{figure}

	\autoref{tab4.1} presents the MARB, DSS and IS values obtained for the missing data in each model.	We summarize the measures for three-years periods with varying percentages of available incidence data. This includes periods 2001-2003 (70\% available), 2004-2006 (75\% available), and so on, up to 2013-2015 (93\% available) (see \autoref{fig4.5} for details). Based on the MARB analysis, Model~3 performs the best for three-year periods with 70\% or 75\%  data availability.
When this percentage increases, all three models show similar MARB performance. For the DSS, Model~3 consistently produces the lowest values, while Model~1 generally yields the highest values. In terms of IS, values are generally high across models, with the lowest values observed when data availability reaches 88\% (2010-2012). Additionally, the IS values exhibit low consistency due to the variability in crude rates, as will be explained below.  Notably, Model~3 achieves the lowest IS values overall.

\begin{figure}[!b]
	\begin{center}
		\includegraphics [width=13.5cm]{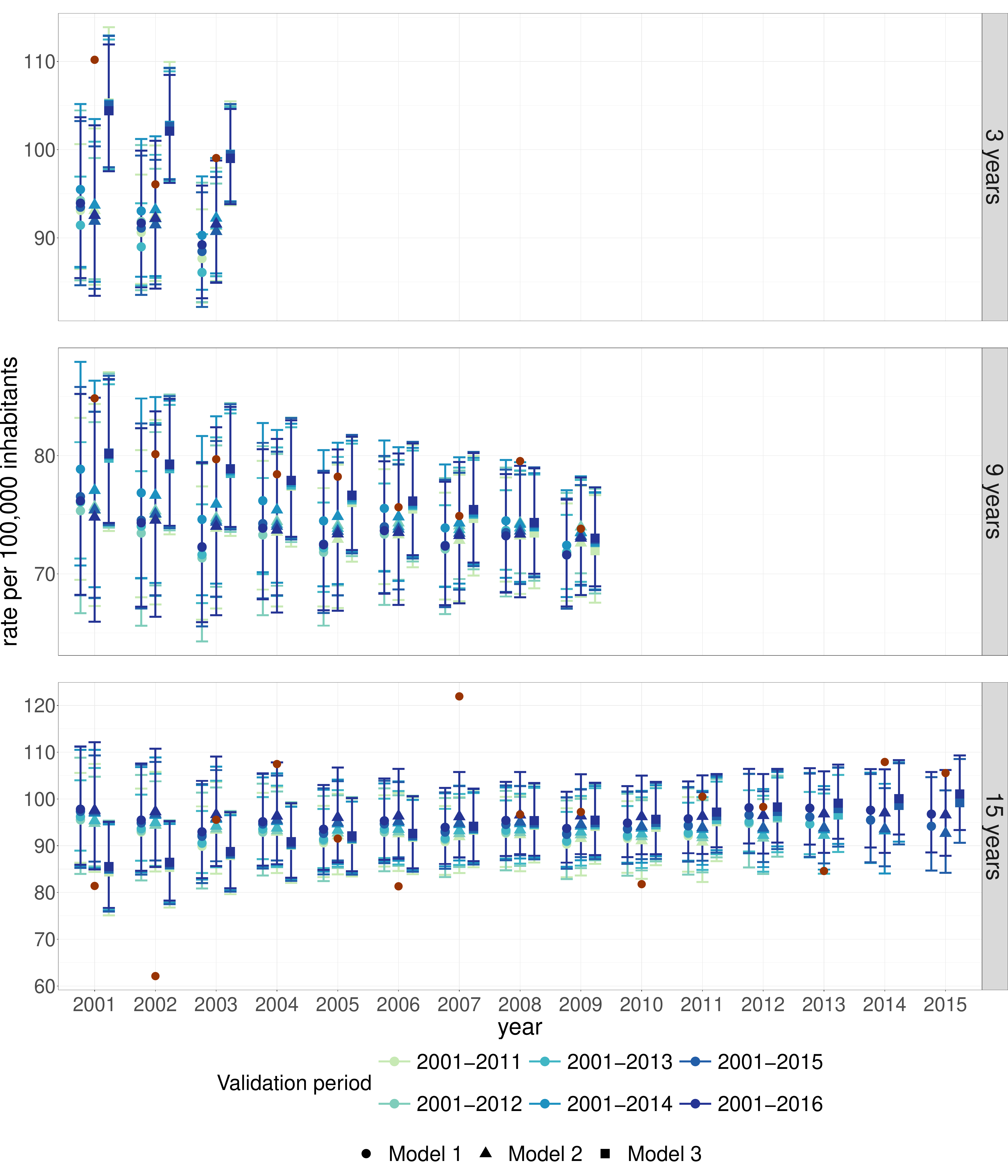}
	\end{center}
	\caption{The posterior median and the 95\% credible interval of the estimated rates per 100,000 inhabitants in years with missing data for three selected areas across the six validation periods defined for the cross-validation procedure. Crude rates are represented by brown dots. Different shapes indicate the models used: dots for Model~1, triangles for Model~2, and squares for Model~3. Colours distinguish the validation periods, ranging from light green for 2001-2011 to dark blue for 2001-2016.\label{fig4.8}}
\end{figure}

\autoref{fig4.9} displays the ARB and DSS average values across validation periods for each group of areas, stratified by the length of the missing data period (3 to 15 years). These groups, visually differentiated by the colour scheme in  \autoref{fig4.5}, span from blue (areas without data in the period 2001–2003) to garnet (areas with 15 years of missing data, from 2001 to 2015).
Results are presented annually to illustrate model performance across different area groups. While no distinct patterns are evident among the area groups due to inherent variability in observed rates, the overall models' performance is favorable. Most ARB values remain below 0.12, except for specific years (2002, 2003, 2004, and 2007) in certain areas.
Unlike ARB, DSS exhibits a more pronounced pattern: areas with the longest data gaps (15 years, garnet) tend to show the lowest DSS values, while areas with 3 and 12 years missing (blue and orange, respectively) have the highest DSS values.
Model~3 generally achieves the lowest ARB values across most area groups, though exceptions occur, such as in 2007 for areas with 12 years missing (orange) and in 2013 for areas with 15 years missing (garnet). Notably, Model~3 consistently maintains the lowest DSS values across all area groups and years. Additional IS results are shown in \autoref{figA4.9} in the \autoref{Appendix:Results}, exhibiting patterns analogous to those identified for the ARB.

\autoref{fig4.8} displays the posterior median and the 95\% credible intervals of the estimated rates per 100,000 inhabitants in years with missing data for three selected areas across the six validation periods of the cross-validation approach (see \autoref{fig4.6} in the \autoref{Appendix:Results} for further details).
The selected areas exhibit varying levels of data completeness: the top panel shows an area with three missing initial years (blue area in \autoref{fig4.5}), the middle panel shows an area with nine missing initial years (pink area in \autoref{fig4.5}), and the bottom panel shows an area with no available data for the initial fifteen years (garnet area in \autoref{fig4.5}). Additionally, \autoref{fig4.8} depicts the observed crude rates (brown dots), with the line width representing increments of five units per 100,000 inhabitants.
\autoref{fig4.8} reveals minor discrepancies in the posterior median and 95\% credible intervals across the six validation periods (represented by different colours), with variations of less than five units per 100,000 inhabitants. Notably, Model~3 exhibits initial disparities in the estimated median compared to Models~1 and 2, which show similar estimates in both median and credible intervals.
Notably, the posterior median values estimated by Model~3 align more closely with observed rates in these early years, indicating greater initial accuracy. As the time series move forward, the estimates obtained by the three models become more similar.	
In certain areas, such as the region with 15 years of missing data shown in \autoref{fig4.8} (bottom panel), high variability in crude rates is observed, occasionally placing observed rates outside the credible intervals.
Unfortunately, cancer incidence data is often subject to this variability due to factors such as incidental findings or overdiagnosis, and to varying levels of completeness due to the multiple sources from which PBCRs compile data \citep{parkin2006, bray2018}.
These variability aligns with the findings in \autoref{fig4.10} in the \autoref{Appendix:Results}, where high ARB values are observed for certain areas, such as area 28, which corresponds to the region depicted in the bottom panel of \autoref{fig4.8}. In summary, the results suggest that Model~3 provides more accurate predictions, particularly for areas with missing data, especially in the initial years where the percentage of missing data is higher.


\subsubsection{Short-term forecasting}

\begin{figure}[!t]
	\begin{center}
		\includegraphics [width=12.75cm]{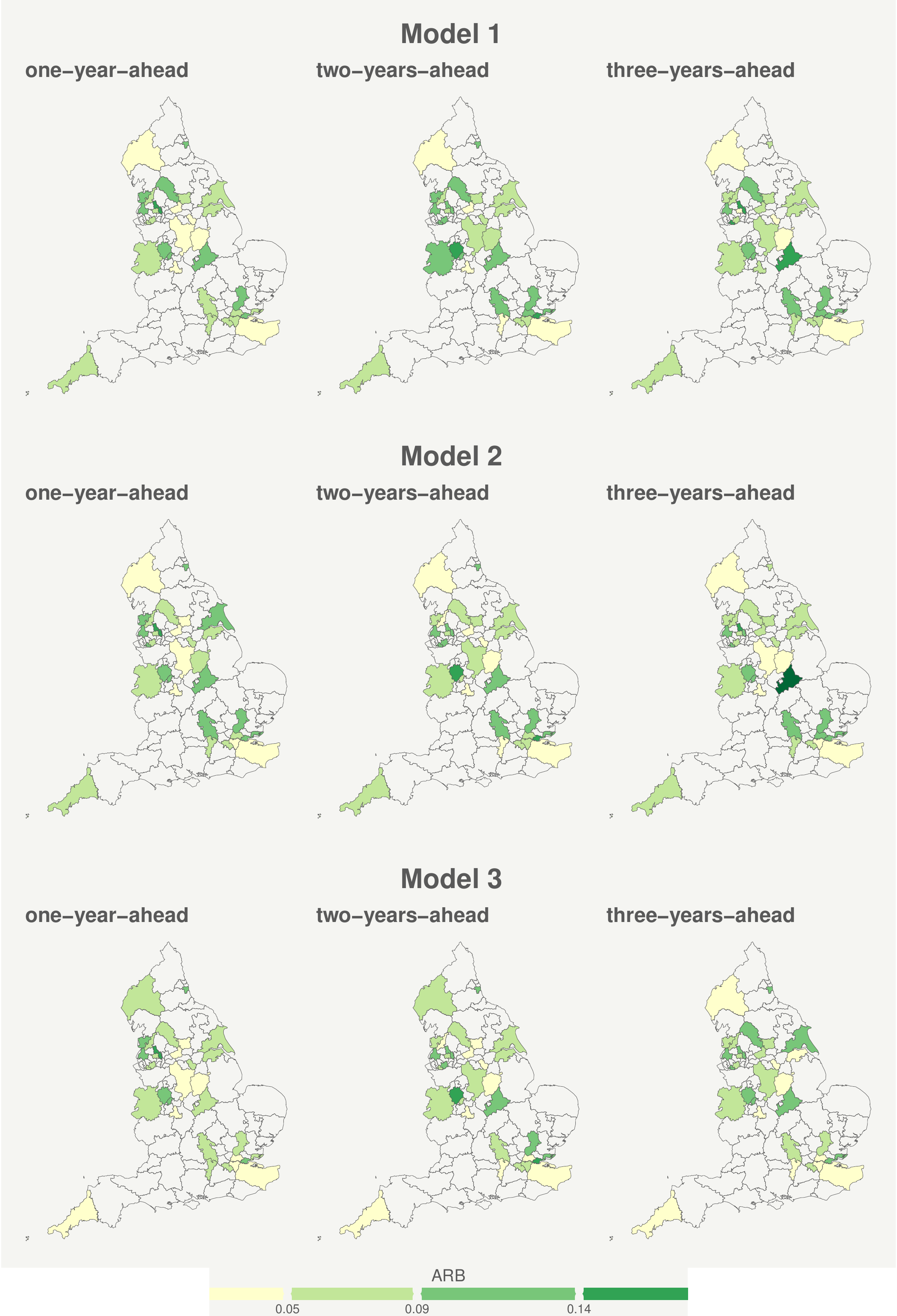}
	\end{center}
	\caption{MARB values for one-, two-, and three-year-ahead predictions are shown for each area with missing data across the time series. The top row displays results using Model~1, the middle row presents results from Model~2, and the bottom row shows the results for Model~3.\label{fig4.11}}
\end{figure}

\autoref{fig4.11} depicts the MARB values across all regions and forecast horizons (years-ahead predictions) for each of the three proposed models, illustrating areas with missing data. Most areas achieve MARB values below 0.09 for all models, reflecting good overall forecasting accuracy. 
Notably, there is no significant deterioration in performance as the forecast horizon increases. However, we note that the areas with the highest and lowest MARB values obtained by each model vary depending on the forecast horizon. This variability is reflected in the distinct spatial distribution of MARB observed for each model across different forecast horizons. Importantly, Model~3 consistently achieves the lowest MARB values in the majority of areas. \autoref{fig4.13} in the \autoref{Appendix:Results} presents results for areas with complete time series. The conclusions remain consistent with those from areas with missing data. Moreover, the MARB values for both groups, areas with missing data and those with complete series are similar, indicating that the number of available years in the series before forecasting has minimal impact on forecasting accuracy. Results for DSS and IS are provided in \autoref{AppendixC}.

\begin{figure}[ !ht]
	\begin{center}
		\includegraphics [width=13.75cm]{ 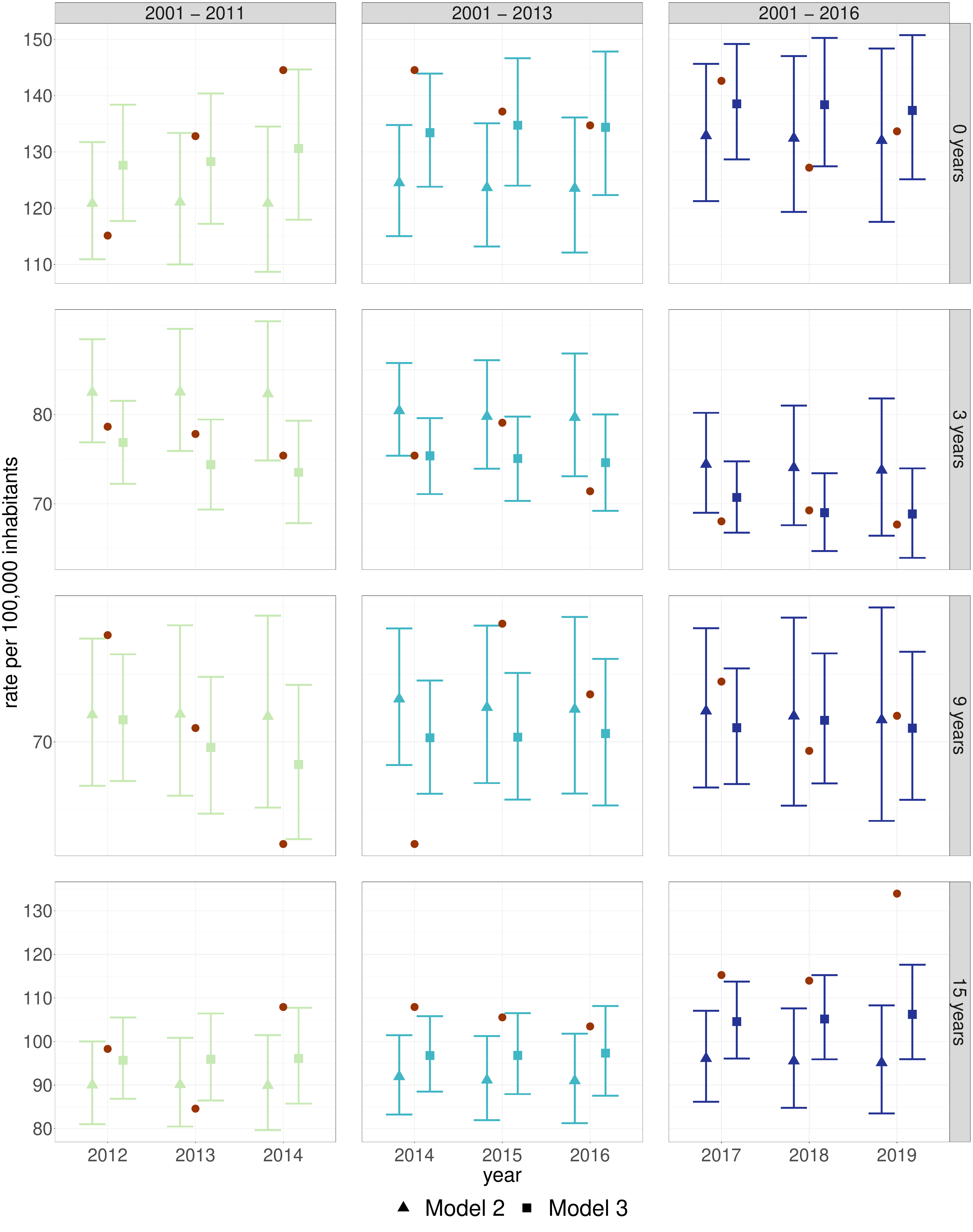}
	\end{center}
	\caption{Posterior median and 95\% credible interval of the forecasted rates per 100,000 inhabitants for four selected areas across validation periods 2001-2011 (left column), 2001-2013 (middle column) and 2001-2016 (right column) obtained by Model~2 (triangles) and Model~3 (squares). Crude rates are represented by brown dots.\label{fig4.12}}
\end{figure}

\autoref{fig4.12} illustrates the posterior median and the 95\% credible interval of the forecasted rates per 100,000 inhabitants for Model~2 and Model~3 in four selected areas across three validation periods defined in the cross-validation approach (see \autoref{fig4.6} in the \autoref{Appendix:Results} for more details). Given the large 95\% credible intervals length (CIL) of Model~1, we exclude it from the analysis. The areas selected for comparison are identical to the three depicted in \autoref{fig4.8}. Additionally, we include a new area where all time series data is available (the lilac area in \autoref{fig4.5}). 
The three validation periods shown correspond to short-term forecasts based on periods from 2001 to 2011, 2001 to 2013 and 2001 to 2016, yielding the three-year forecasts 2012-2014, 2014-2016 and 2017-2019, respectively.
The forecasted rates for remaining validation periods are depicted in \autoref{AppendixC}. The crude rates are represented by brown dots in \autoref{fig4.12}. Disparities in the posterior estimates are observed between Model~2 and Model~3.
Interestingly, predictions from Model~2 for each validation period are nearly constant across all forecast horizons. In contrast, forecasted values of Model~3 are more variable. For example, forecasted rates for 2012-2014 show contrasting trends: a clear increase in areas with no missing years and a decrease in areas with three missing years. In general, Model~3 provides more accurate predictions.

\section{National Incidence}\label{Section:National}
A key objective in analyzing cancer burden is obtaining accurate estimates of national cancer incidence. INLA does not directly provide the posterior distribution of the counts $O_{itI}$ for each area $i$ and time $t$. Consequently, it does not offer the national incidence for each year. Therefore, additional programming within \texttt{R-INLA} is required to derive the national incidence. Similar to \cite{retegui2023}, we use a four-step sampling technique to compute the posterior empirical distribution of national incidence for each year. First, we generate $S$ samples from the joint posterior distribution of the linear predictor and hyperparameters for each area $i$ and time $t$ using \texttt{inla.posterior.sample()}, obtaining samples on the log scale. These samples are then transformed to the observation scale ($r_{itI}$) using \texttt{inla.posterior.sample.eval()} with the exponential function. Next, we sample counts from a Poisson distribution with mean $\mu_{itI} = n_{itI}r_{itI}$ for each area and time, producing $S$ samples of counts, denoted $O^s_{itI}$ where $s = 1 \dots, S$.  These samples could be organized into $T$ tables, one for each year $t$, where each row represents the sample counts for each area $i$. An example of this structure is provided in \autoref{tab2.1} in the \autoref{Appendix:Results}. Finally, we derive the posterior empirical distribution of national incidence by summing the counts across areas for each sample $s$, meaning we do a columnwise sum of \autoref{tab2.1} in the \autoref{Appendix:Results}, resulting in $S$ samples of national incidence for each year.

\begin{table}[ !ht]
	\centering
	\caption{\label{tab4.4} Observed national incidence counts, posterior mean of estimated national incidence counts and the 95\% credible interval estimated by each model for the validation period from 2001 to 2013.}
			\resizebox{0.75\textwidth}{!}{
		\begin{tabular}{rc|crcrc}
			\toprule
			\multicolumn{2}{c}{ }&\bf Model 1&&\bf Model 2&&\bf Model 3\\
			\cline{3-3}\cline{5-5}\cline{7-7}
			&&&&&&\\[-0.8em]
			&\bf Observed &\bf Counts&&\bf Counts&&\bf Counts\\
			\midrule
			2001 & 18,948 & 18,831 &  & 18,575 &  & 18,456 \\
			&&(18,405  -  19,268) &  & (18,216  -  18,950) &  & (18,130  -  18,788) \\
			&&&&&&\\[-0.4em]
			2002 & 18,258 & 18,414 &  & 18,557 &  & 18,453 \\
			&&(17,995  -  18,849) &  & (18,227  -  18,896) &  & (18,137  -  18,765) \\
			&&&&&&\\[-0.4em]
			2003 & 17,948 & 17,900 &  & 18,477 &  & 18,440 \\
			&&(17,487  -  18,325) &  & (18,168  -  18,794) &  & (18,136  -  18,747) \\
			&&&&&&\\[-0.4em]
			2004 & 18,470 & 18,590 &  & 18,505 &  & 18,502 \\
			&&(18,178  -  18,998) &  & (18,196  -  18,812) &  & (18,202  -  18,809) \\
			&&&&&&\\[-0.4em]
			2005 & 18,280 & 18,336 &  & 18,617 &  & 18,650 \\
			&&(17,947  -  18,735) &  & (18,313  -  18,925) &  & (18,360  -  18,949) \\
			&&&&&&\\[-0.4em]
			2006 & 18,946 & 18,932 &  & 18,835 &  & 18,807 \\
			&& (18,530  -  19,335) &  & (18,527  -  19,150) &  & (18,506  -  19,115) \\
			&&&&&&\\[-0.4em]
			2007 & 18,786 & 18,692 &  & 18,932 &  & 18,918 \\
			&&(18,295  -  19,100) &  & (18,623  -  19,243) &  & (18,610  -  19,223) \\
			&&&&&&\\[-0.4em]
			2008 & 19,237 & 19,240 &  & 19,195 &  & 19,124 \\
			&& (18,843  -  19,640) &  & (18,889  -  19,507) &  & (18,834  -  19,427) \\
			&&&&&&\\[-0.4em]
			2009 & 18,978 & 18,944 &  & 19,302 &  & 19,286 \\
			&&(18,540  -  19,346) &  & (18,991  -  19,608) &  & (18,987  -  19,587) \\
			&&&&&&\\[-0.4em]
			2010 & 19,360 & 19,356 &  & 19,449 &  & 19,493 \\
			&&(18,961  -  19,755) &  & (19,131  -  19,764) &  & (19,174  -  19,813) \\
			&&&&&&\\[-0.4em]
			2011 & 19,738 & 19,660 &  & 19,655 &  & 19,728 \\
			&&(19,254  -  20,069) &  & (19,343  -  19,967) &  & (19,427  -  20,038) \\
			&&&&&&\\[-0.4em]
			2012 & 20,321 & 20,317 &  & 19,867 &  & 19,979 \\
			&&(19,913  -  20,735) &  & (19,559  -  20,187) &  & (19,672  -  20,303) \\
			&&&&&&\\[-0.4em]
			2013 & 20,385 & 20,394 &  & 20,079 &  & 20,177 \\
			&&(20,006  -  20,784) &  & (19,753  -  20,406) &  & (19,852  -  20,505) \\
			\bottomrule
		\end{tabular}
					}
\end{table}

\autoref{tab4.4} presents the observed national incidence counts alongside the estimated mean national incidence counts and 95\% credible intervals for each model during the validation period from 2001 to 2013. In most years, the estimated mean national incidence values are similar across the three models; however, the largest discrepancies appear in 2001 and 2003, where Model~1 yields more accurate mean estimates. However, Model~1 also produces the widest credible intervals, with lengths ranging from 750 to 870. These findings are consistent across all validation periods, as shown in \autoref{fig4.14} in the \autoref{Appendix:Results}, which depicts the ARB and 95\% CIL values obtained by each proposed model for the years of each validation period. Results indicate that all models yield national incidence estimates with low relative bias, typically below 2\%. While Model 1 exhibits slightly better overall bias performance, it also has the widest confidence intervals. 

For short-term national incidence forecasts, \autoref{tab4.5} displays MARB and 95\% CIL values for one-, two- and three-year-ahead predictions. Model~1 exhibits increasing MARB values as the prediction horizon extends. In contrast, Model~2 shows a decrease in MARB with longer horizons, while Model~3 maintains stable MARB values, averaging around 1.5\% across all forecast periods. Notably, Model~3 achieves the lowest MARB values particularly for three years ahead predictions, suggesting its potential for superior short-term forecasting accuracy. As expected, the CIL widens as the prediction horizon increases. In this regard, Model~1 exhibits the widest credible intervals, indicating greater uncertainty in its short-term forecasts, while Model~3 demonstrates the narrowest intervals, suggesting higher confidence in its short-term predictions. Additionally \autoref{tab4.6} provides the observed national incidence counts, the projected mean national incidence counts and 95\% credible intervals for each model during the short-term forecasting for the validation period from 2001 to 2013. Consistent with the results summarized in \autoref{tab4.5}, Model~3 produces the most accurate projections, while Model~1 is associated with the largest credible intervals.

\begin{table}[ !ht]
	\centering
	\caption{\label{tab4.5} MARB and 95\% CIL values obtained by each model for the short-term forecasting of national incidence.}
	\resizebox{0.82\textwidth}{!}{
	\begin{tabular}{rc|ccrccrcc}
		\toprule
		\multicolumn{2}{c}{ }&\multicolumn{2}{c}{\bf Model 1}&&\multicolumn{2}{c}{\bf Model 2}&&\multicolumn{2}{c}{\bf Model 3}\\
		\cline{3-4}\cline{6-7}\cline{9-10}
		&&&&&&&&&\\[-0.8em]
		\multicolumn{2}{l}{\bf Predicted years}&\bf MARB&\bf CIL&&\bf MARB&\bf CIL&&\bf MARB&\bf CIL\\
		\midrule
		&&&&&&&&&\\[-0.8em]
		\multicolumn{2}{l|}{one-year-ahead} &0.015 & 4,522.00  &  & 0.020 & 760.833  &  & 0.015 & 708.833 \\
		&&&&&&&&&\\[-1em]
		\multicolumn{2}{l|}{two-years-ahead} &0.016 & 6,399.50  &  & 0.018 & 825.000 &  & 0.015 & 760.833  \\
		&&&&&&&&&\\[-1em]
		\multicolumn{2}{l|}{three-years-ahead} &0.020 & 8,072.50  &  & 0.016 & 904.333  &  & 0.014 & 802.500 \\
		\bottomrule
	\end{tabular}
}
\end{table}

\begin{table}[ t]
	\centering
	\caption{\label{tab4.6} Observed national incidence counts, posterior mean of estimated national incidence counts and the 95\% credible interval for short-term projections from each model during the validation period from 2001 to 2013.}
	\resizebox{0.82\textwidth}{!}{
		\begin{tabular}{rc|crcrc}
			\toprule
			\multicolumn{2}{c}{ }&\bf Model 1&&\bf Model 2&&\bf Model 3\\
			\cline{3-3}\cline{5-5}\cline{7-7}
			&&&&&&\\[-0.8em]
			&\bf Observed &\bf Counts&&\bf Counts&&\bf Counts\\
			\midrule
			2014 & 20,474 & 20,642 &  & 20,211 &  & 20,349 \\
			&& (17,625  -  23,891) &  & (19,844  -  20,568) &  & (19,999  -  20,702) \\
			&&&&&&\\[-0.4em]
			2015 &  20,416 & 20,983 &  & 20,257 &  & 20,557 \\
			&&(16,742  -  25,776) &  & (19,833  -  20,649) &  & (20,181  -  20,937) \\
			&&&&&&\\[-0.4em]
			2016 & 21,036  & 21,335 &  & 20,473 &  & 20,800 \\
			&&(16,184  -  27,186) &  & (20,015  -  20,893) &  & (20,413  -  21,206) \\ 				
			\bottomrule
		\end{tabular}
		}
\end{table}

\newpage
\section{Discussion}
This work aims to achieve two primary objectives: first, to evaluate the effectiveness of multivariate spatio-temporal models in completing cancer registry data series; and second, to use these models for short-term cancer incidence forecasting, ultimately enabling accurate national cancer incidence projections.
Specifically, three different models have been proposed: one model with a shared spatial component plus disease specific intercept, temporal and interaction terms (Model~1); a second integrating both spatial and temporal shared components (Model~2); and a third model incorporating a spatial shared component alongside a flexible shared spatio-temporal component (Model~3). The flexible shared spatio-temporal component incorporates a time-varying scaling parameter. This allows the model to adjust the space-time interactions between cancer incidence and mortality data along the time.
To evaluate the predictive performance of the models, we conducted a validation study using lung cancer incidence and mortality data for clinical commissioning groups in England from 2001 to 2019. This design aimed to replicate real-world scenarios documented in the literature.

Results for missing data indicate that in years with larger number of missing data, specifically when 70\% to 80\% of the incidence data is available, the model featuring both spatial shared components and flexible shared spatio-temporal effects (Model~3) performs the best, producing predictions that more closely resemble the observed crude rates. However, as the percentage of available incidence data increases, all models become equally competitive. High IS values are observed for all models likely stem from the large variability of rates observed in certain areas, as previously discussed.
This variability is not fully captured by the models, whose primary objective is to smooth the rates. 
Regarding the results for short-term forecasting, we observe that the number of available years in the time series prior to forecasting does not significantly impact the accuracy of the projections. Moreover, similar results in terms of bias are observed for one-, two- and three-years-ahead predictions. However, an increase in IS is noted for Model~1 as the number of years-ahead predictions increases. Model~3 presents the lowest values for all criteria in most of the areas. Mirroring the performance with missing data, Model~3 yields the most accurate forecasts. Notably, this model captures varying trends in the predicted years, unlike other models that predict a nearly constant rate across the entire forecast horizon. Finally, national incidence estimates are provided for each year in the time series. Results show that all models present low relative bias, generally around 1\%, with Model~1 achieving the most accurate estimates overall. For short-term forecasting, Model~3 consistently demonstrates the lowest bias and narrowest credible intervals, achieving biases of 1.5\%, 1.5\%, and 1.4\% for one-, two-, and three-year-ahead predictions, respectively. In summary, the findings from this study support recommending the model featuring both spatial shared components and flexible shared spatio-temporal effects (Model~3) for completing time series data and generating short-term projections at the regional level. We would also recommend Model~3 for short-term forecasting of national incidence, although Model~1 yields accurate results when estimating national incidence counts.

To conclude, it is important to note that we did some additional analyses using colorectal cancer data for both sexes, revealing that Model~2 consistently produced the least accurate results, exhibiting inconsistencies across validation periods.  In contrast, Models~1 and 3 performed similarly, with Model~3 showing slight improvements in certain areas. Additionally, when analyzing total cancer data for males and females, all three models demonstrated comparable performance, suggesting that for cancers with high incidence and mortality rates, the proposed models behave similarly. The findings from this work provide valuable insights for health authorities and policymakers.  By providing annual estimates of cancer incidence at both national and regional levels, this research provides decision-makers with critical information to inform health care policies and resource allocation.

\section*{Declaration of interests}
All authors certify that they have no affiliations with or involvement in any organization
or entity with any financial interest or non-financial interest in the subject matter or materials discussed in this manuscript.

\section*{Funding sources}
The work was supported by Project PID2020-113125RB-I00/MCIN/AEI/10.13039/501100011033, Ayudas Predoctorales Santander UPNA 2021–2022 and Research Networks Project BIOSTATNET AEI/RED2024-153680-T.

\bibliography{BIB_October2024}

\newpage
\begin{appendices}
	
	\section{Additional Figures and Tables}\label{Appendix:Results}
	
	\begin{figure}[!ht]
		\begin{center}
			\includegraphics[width=16cm,  page=1]{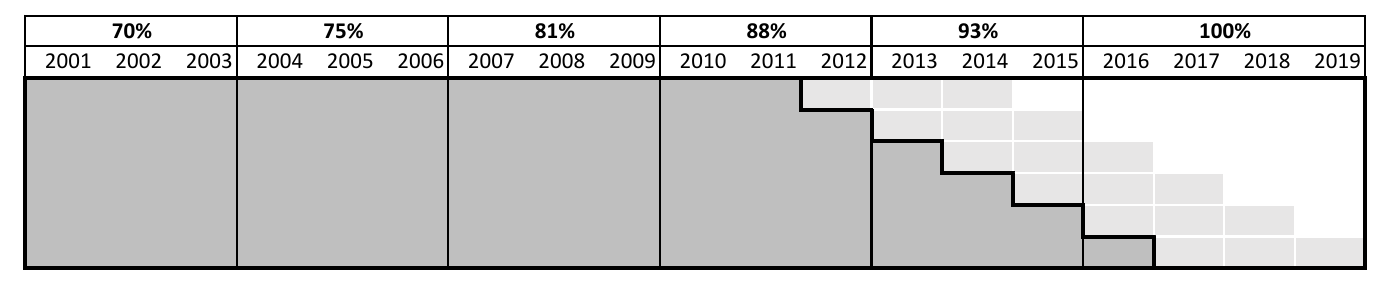}
		\end{center}
		\caption{The figure depicts the distribution of data across the $K=6$ periods within the cross-validation framework. Grey bars indicate the observed period for each validation period, whereas light grey bars denote the periods to be forecasted. The percentages displayed at the top of the figure represent the percentage of areas with available incidence data during each three-year period.\label{fig4.6}}
	\end{figure}
	
	\begin{figure}[ !h]
		\begin{center}			
			\includegraphics [width=15cm, page=3]{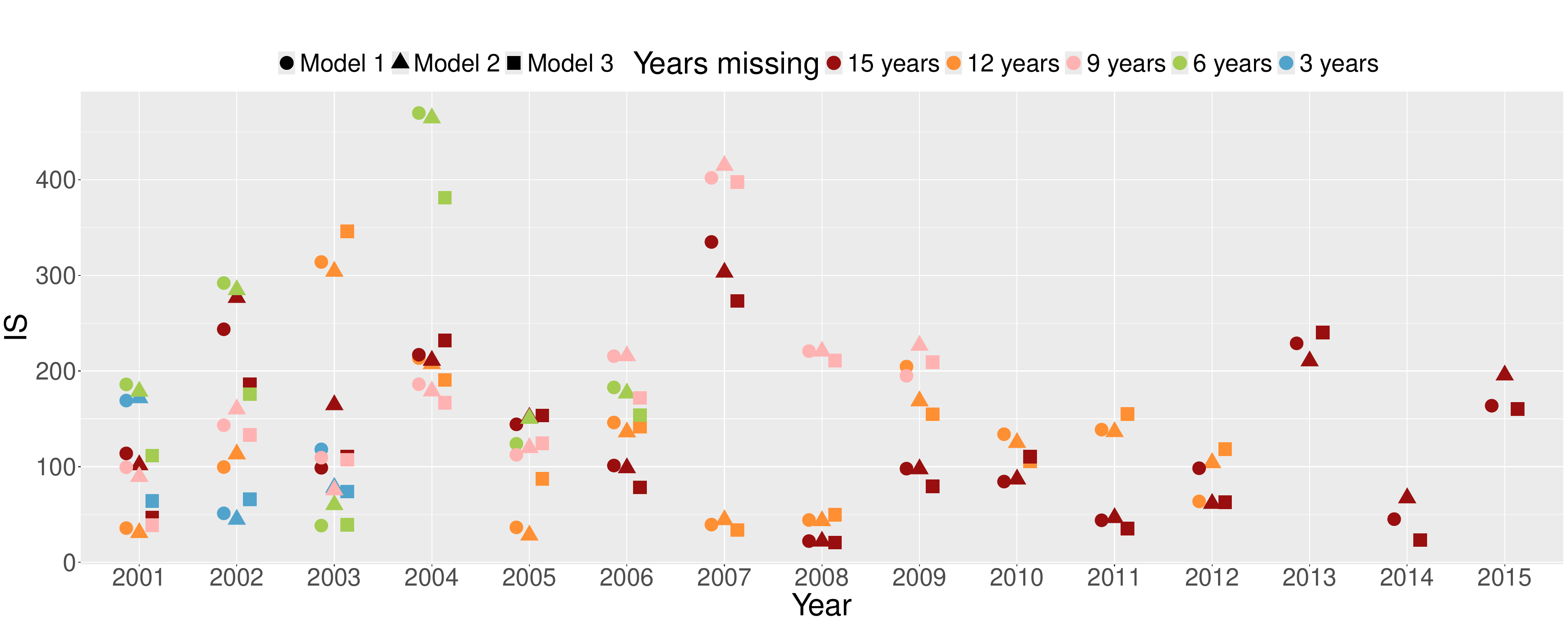}
		\end{center}
		\caption{Average values across validation periods for IS, represented for each year of the time series and each group of areas based on the duration of missing data. Different shapes indicate the models used: dots for Model~1, triangles for Model~2, and squares for Model~3. Colours represent different groups of areas based on missing data duration: blue for areas with 3 years missing, green for 6 years, pink for 9 years, orange for 12 years, and garnet for 15 years. \label{figA4.9} }
	\end{figure}
	
	\begin{figure}[ !h]
		\begin{center}
			\includegraphics [width=13.5cm, page=3]{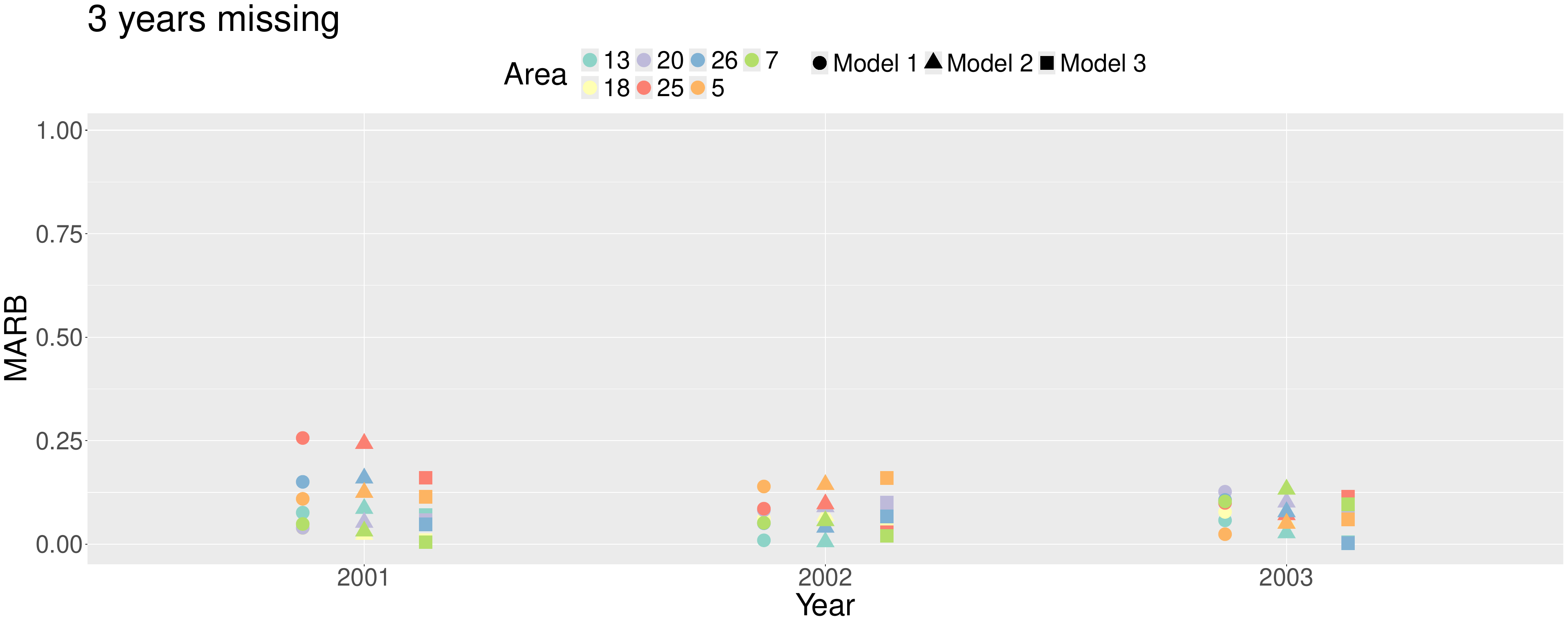}\\ \vspace*{0.5cm}
			
			\includegraphics [width=13.5cm, page=5]{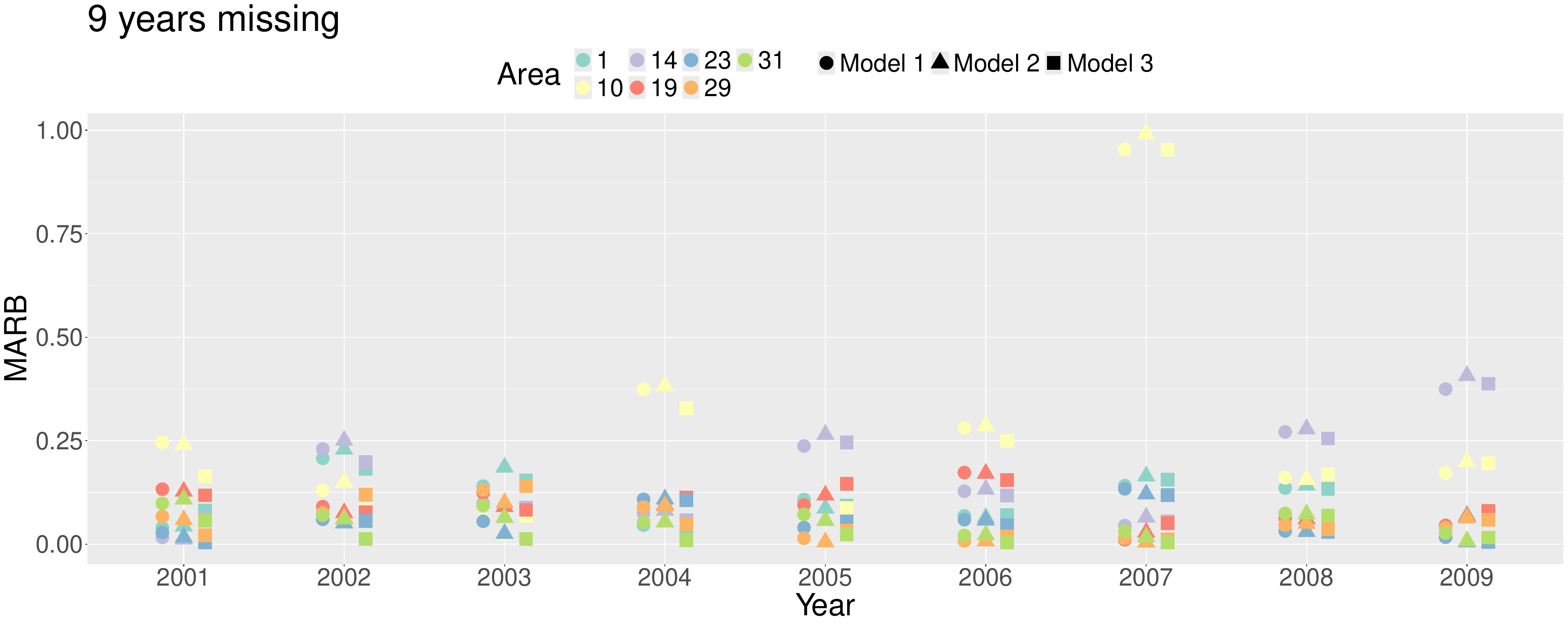}\\ \vspace*{0.5cm}
			
			\includegraphics [width=13.5cm, page=2]{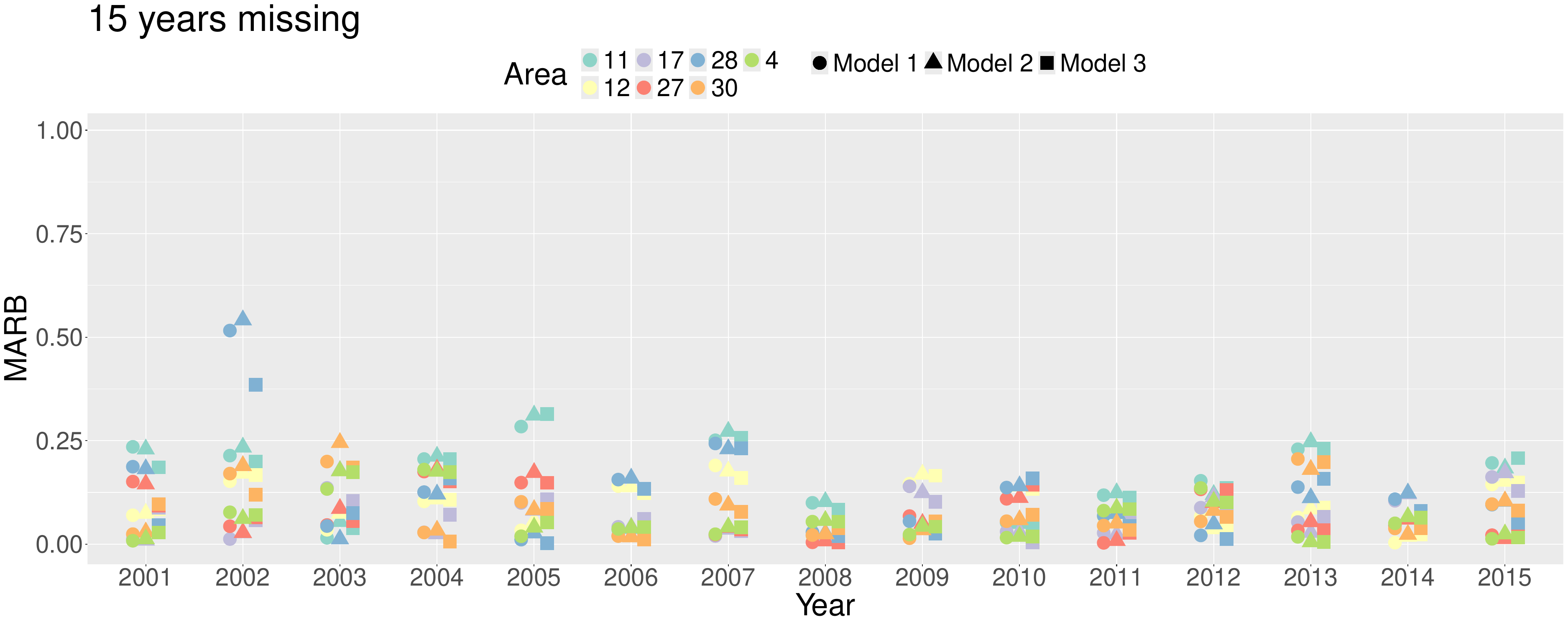}
		\end{center}
		\caption{Average values across validation periods for ARB achieved by each model,  shown for areas with 3 years missing data (top panel), 9 years missing (middle panel), and 15 years missing (bottom panel). Different shapes indicate the models used: dots for Model~1, triangles for Model~2, and squares for Model~3. Colours distinguish different areas within each group. \label{fig4.10} }
	\end{figure}
	
	\begin{figure}[ !h]
		\begin{center}
			\includegraphics [width=12.25cm]{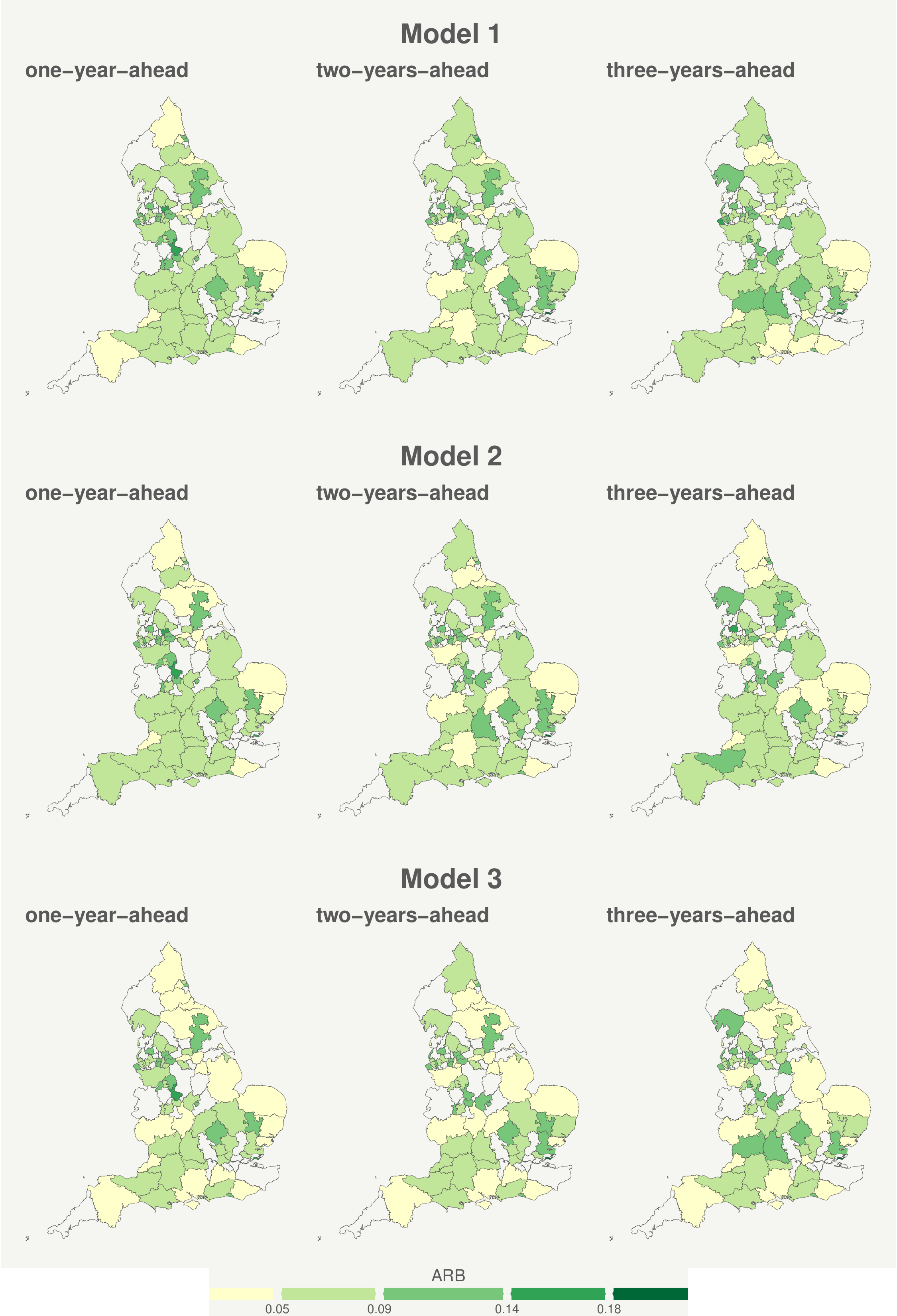}
		\end{center}
		\caption{MARB values for one-, two-, and three-year-ahead predictions are shown for each area with no missing data across the time series. The top row displays results using Model~1, the middle row presents results from Model~2, and the bottom row shows the results for Model~3.\label{fig4.13}}
	\end{figure}
	
	\clearpage
	
	\begin{table}[ !ht]
		\centering
		\caption{Visualization of sample organization from area-specific counts at time $t$. \label{tab2.1}}
		\begin{tabular}{c|cccc}
			& Sample 1 & Sample 2 & \dots & Sample $S$\\
			\hline
			&&&&\\[-0.8em]
			Area 1 &$O^1_{1tI}$ & $O^2_{1tI}$ & \dots & $O^S_{1tI}$\\
			&&&&\\[-1em]
			Area 2&$O^1_{2tI}$ & $O^2_{2tI}$ & \dots & $O^S_{2tI}$\\
			&&&&\\[-1em]
			\vdots&\vdots & \vdots & \dots &\vdots\\
			&&&&\\[-1em]
			Area $A$&$O^1_{AtI}$ & $O^2_{AtI}$ & \dots & $O^S_{AtI}$\\
			\multicolumn{2}{l}{ }&&&\\[-1em]
			\hline
		\end{tabular}
	\end{table}
	
	\vspace{2cm}
	
	\begin{figure}[!h]
		\begin{center}
			\includegraphics [width=14cm]{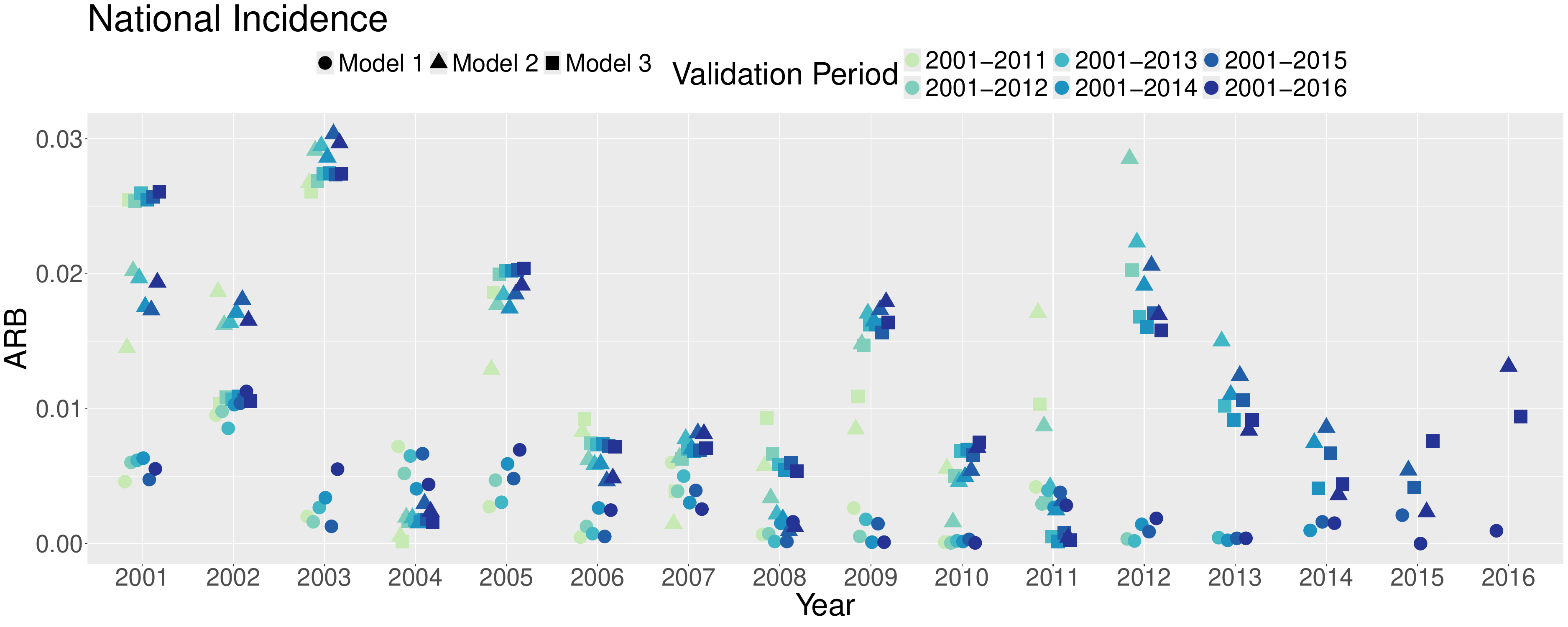} \\ \vspace*{0.3cm}
			
			\includegraphics [width=14cm]{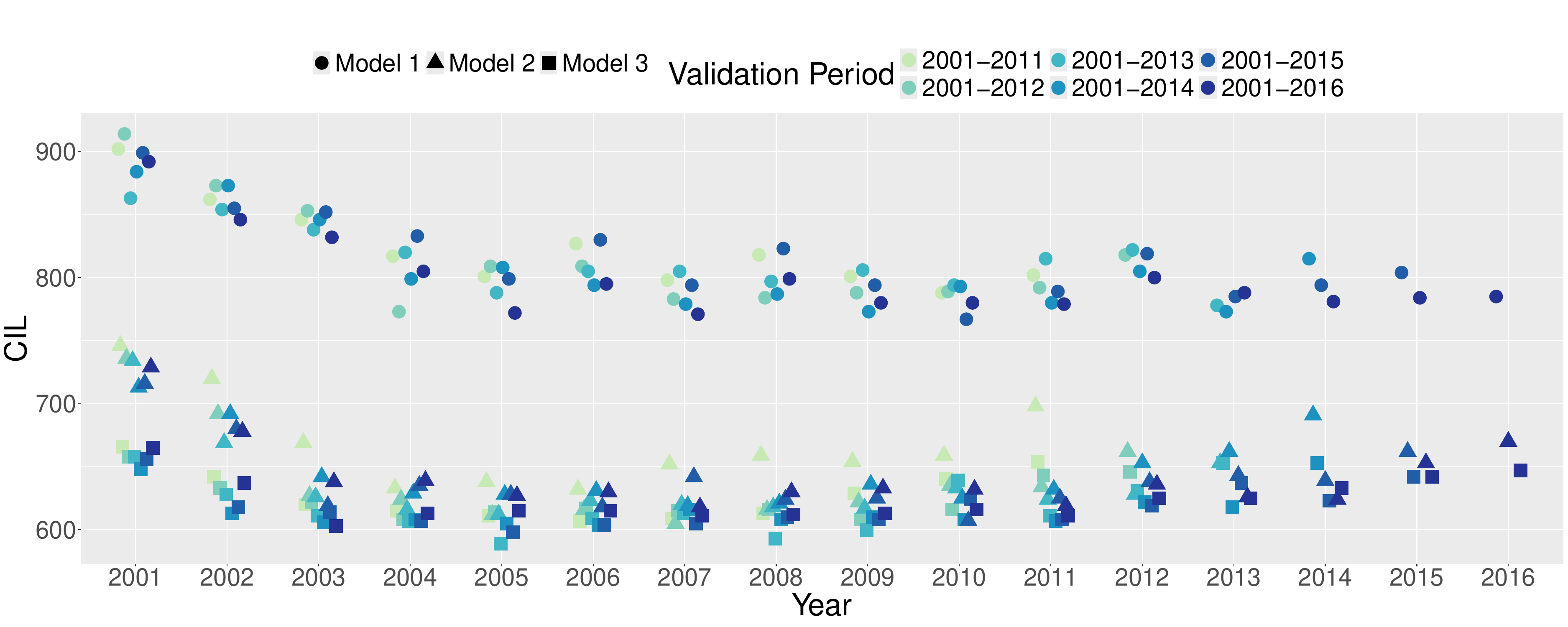}
		\end{center}
		\caption{ARB values and 95\% credible interval length (CIL) obtained by Model~1 (dots), Model~2 (triangles) and Model~3 (squares) are shown for each year across the validation periods. Colours distinguish the validation periods, ranging from light green for 2001-2011 to dark blue for 2001-2016. \label{fig4.14} }
	\end{figure}
	
	\section{Summary of hyperparameters}\label{AppendixA}
	This section presents a summary and discussion of the main hyperparameters from the models introduced in the main paper. Since our analysis includes six validation periods, we obtain six posterior distributions for each hyperparameter. To evaluate the consistency of these estimates across periods and provide a concise summary, \autoref{tab5} reports the median, minimum, and maximum of the posterior medians for each hyperparameter over all six validation periods. 
	Among all components, the unstructured random effect $u$ consistently shows the highest precision values, with notable variation between its minimum and maximum. This indicates that in at least one validation period, its estimate may be less reliable. Nevertheless, despite its variability, excluding this effect from the model leads to poorer predictive performance. For this reason, we retain it in the analysis. In contrast, the estimates for the remaining hyperparameters demonstrate greater stability across validation periods.
	
	\begin{table}[h!]
		\centering
		\caption{Summary of the posterior median of the hyperparameters across the six validation periods.\label{tab5}}
		\resizebox{\textwidth}{!}{
			\begin{tabular}{l|rrrlrrrlrrr}
				\hline
				\multicolumn{1}{c}{ }&\multicolumn{3}{c}{\textbf{Model 1}}&&\multicolumn{3}{c}{\textbf{Model 2}}&&\multicolumn{3}{c}{\textbf{Model 3}}\\
				\cline{2-4}\cline{6-8}\cline{10-12}
				& min & 50\% & max && min & 50\% & max && min & 50\% & max\\ 
				\hline
				$\tau_\kappa$ & 24.15 &  35.15 & 39.46 && 27.74 & 33.16  & 36.90 &&  24.87 & 30.14 & 39.11 \\ 
				$\delta$  & 1.00 & 1.01 &  1.02 && 1.00 & 1.00 &  1.02 && 1.00 & 1.00 & 1.01\\ 
				$\tau_{u}$  & 1606.39 &  19908.54 &746252.91 && 1746.30 & 3764.44 &  582487.74 &&  2282.95 & 4377.61 &37124.86\\ 
				&&&\\
				$\tau_{\gamma_I}$  & 108.37 & 372.14 & 984.53 \\ 
				$\tau_{\gamma_M}$  & 146.89 & 772.80 &  1789.14 &&&&&& 135.53 & 178.74 & 279.96\\ 
				$\tau_{\gamma}$&&&&&  143.73 & 276.67 & 1012.45 \\ 
				$\varsigma$ &&&&&  0.49 & 0.55 & 0.66 \\ 
				&&&\\
				$\tau_{\chi_I}$  & 1340.56 & 1612.21 &  7315.57  && 1338.00 & 1799.80 &  4320.78 \\ 
				$\tau_{\chi_M}$ & 1445.84 & 1696.41 &  1877.99 && 1320.40 & 1691.39 &  2853.61 \\ 
				$\tau_{\chi}$ &&&&&&&&& 1070.99 &  1149.11 &  1316.46 \\ 
				$\varrho$ &&&&&&&&& 0.99 & 1.02 &  1.05 \\ 
				\hline
		\end{tabular} }
	\end{table}

	\section{Predicting area-specific incomplete time series}\label{AppendixB}
	
	\begin{figure}[t!]
		\begin{center}
			\includegraphics [width=14cm, page=4]{Fig_SUP/Figure8.pdf}\\ \vspace*{0.5cm}
			
			\includegraphics [width=14cm, page=1]{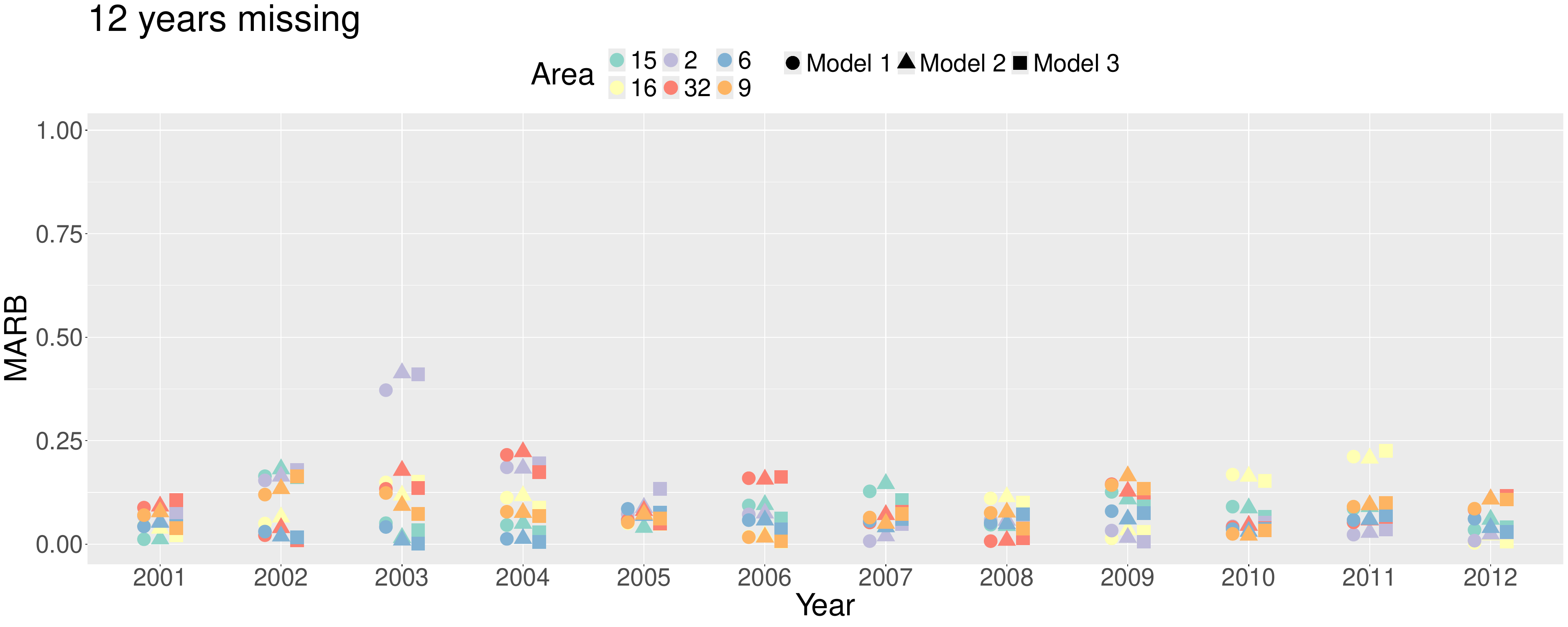}
		\end{center}
		\caption{Average values across validation periods for ARB achieved by each model,  shown for areas with 6 years missing data (top panel) and 12 years missing (bottom panel). Different shapes indicate the models used: dots for Model~1, triangles for Model~2, and squares for Model~3. Colours distinguish different areas within each group. \label{figA1} }
	\end{figure}
	
	This section presents and discusses figures not included in the main paper. Specifically, it compiles the ARB results for areas with 6 and 12 years of missing data in the time series, as well as the DSS and IS results for areas with 3, 6, 9, 12, and 15 years of missing data.

	\begin{figure}[b!]
		\begin{center}
			\includegraphics [width=14cm, page=3]{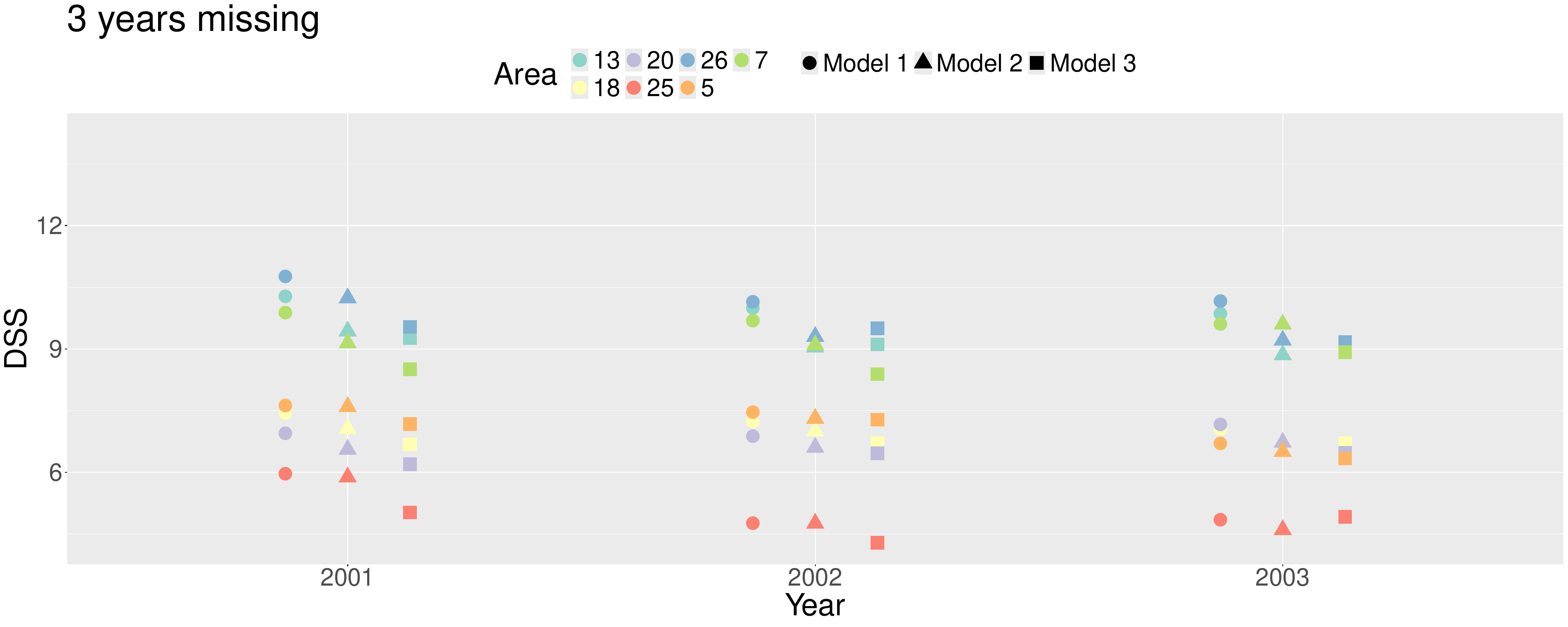}\\ \vspace*{0.5cm}
			
			\includegraphics [width=14cm, page=4]{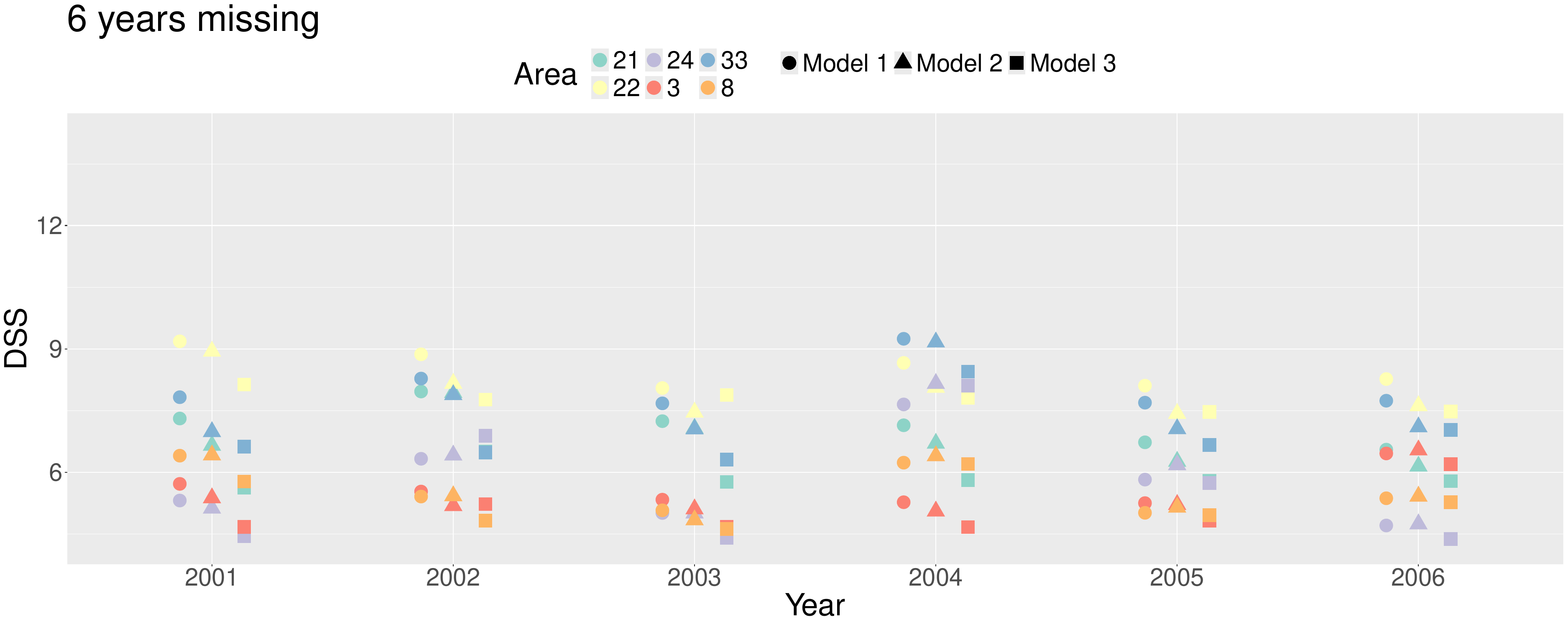}\\ \vspace*{0.5cm}
			
			\includegraphics [width=14cm, page=5]{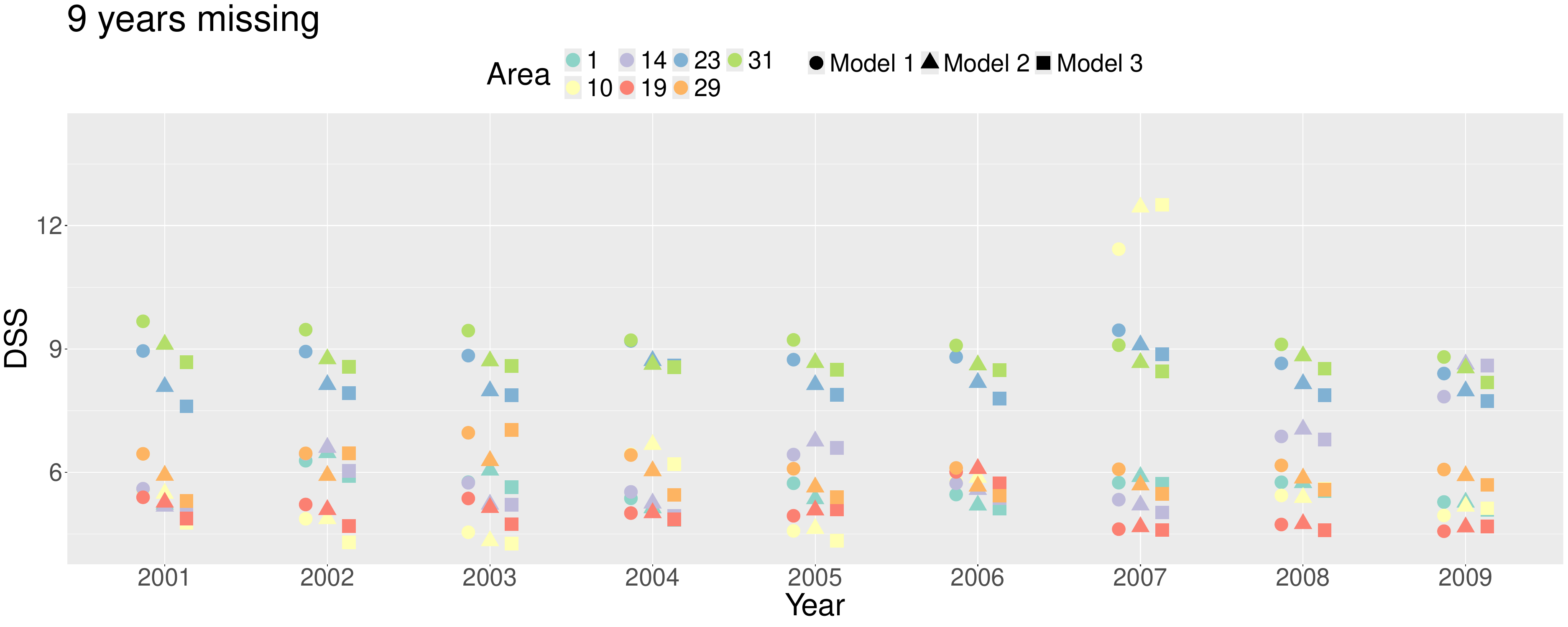}
		\end{center}
		\caption{Average values across validation periods for DSS achieved by each model,  shown for areas with 3 years missing data (top panel), 6 years missing (middle panel), and 9 years missing (bottom panel). Different shapes indicate the models used: dots for Model~1, triangles for Model~2, and squares for Model~3. Colours distinguish different areas within each group. \label{figA2.a} }
	\end{figure}
	
	\autoref{figA1} shows the average ARB values across validation periods achieved by each model, focusing on areas not included in the main paper. Specifically, it highlights regions with 6 years of missing data (green areas in Figure~5 of the main paper) and 12 years of missing (orange areas). Similar to Figure~13 in the Appendix of the main paper, no clear spatial pattern emerges among the areas. In general, ARB values are lower than 0.25, with a few exceptions due to the variability in crude rates observed in certain regions, as discussed in the main paper. While Model~3 performs better in the earlier years, all models exhibit similar behaviour as the time series progresses.

	\begin{figure}[b!]
		\begin{center}
			\includegraphics [width=14cm, page=1]{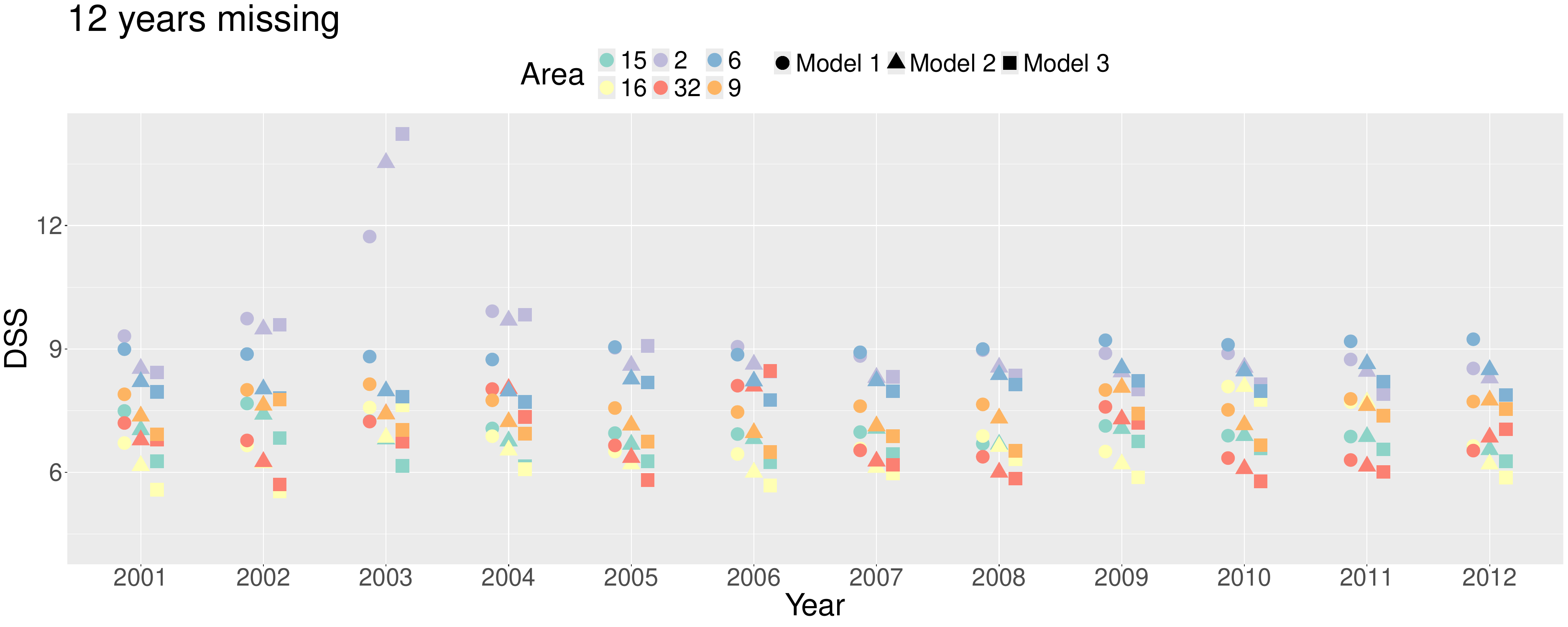}\\ \vspace*{0.5cm}
			
			\includegraphics [width=14cm, page=2]{Fig_SUP/FigureA2.pdf}
		\end{center}
		\caption{Average values across validation periods for DSS achieved by each model,  shown for areas with 12 years missing data (top panel) and 15 years missing (bottom panel). Different shapes indicate the models used: dots for Model~1, triangles for Model~2, and squares for Model~3. Colours distinguish different areas within each group. \label{figA2.b} }
	\end{figure}
	
	\autoref{figA2.a} and \autoref{figA2.b} display the average values across validation periods for DSS, achieved by each model and focusing on areas with 3 years missing data (blue areas in Figure~5 of main paper), 6 years missing (green areas) and 9 years missing (pink areas), and 12 years missing (orange areas) and 15 years missing (garnet areas), respectively. Unlike the ARB results, a discernible pattern emerges among the areas. Generally, areas with high DSS values relative to others tend to maintain these high values throughout the time series, while areas with low or medium values exhibit similar consistency. Additionally, Model~3 consistently achieves the lowest DSS values compared to Model~1 and Model~2, highlighting its superior performance.

	\begin{figure}[t!]
		\begin{center}
			\includegraphics [width=14cm, page=3]{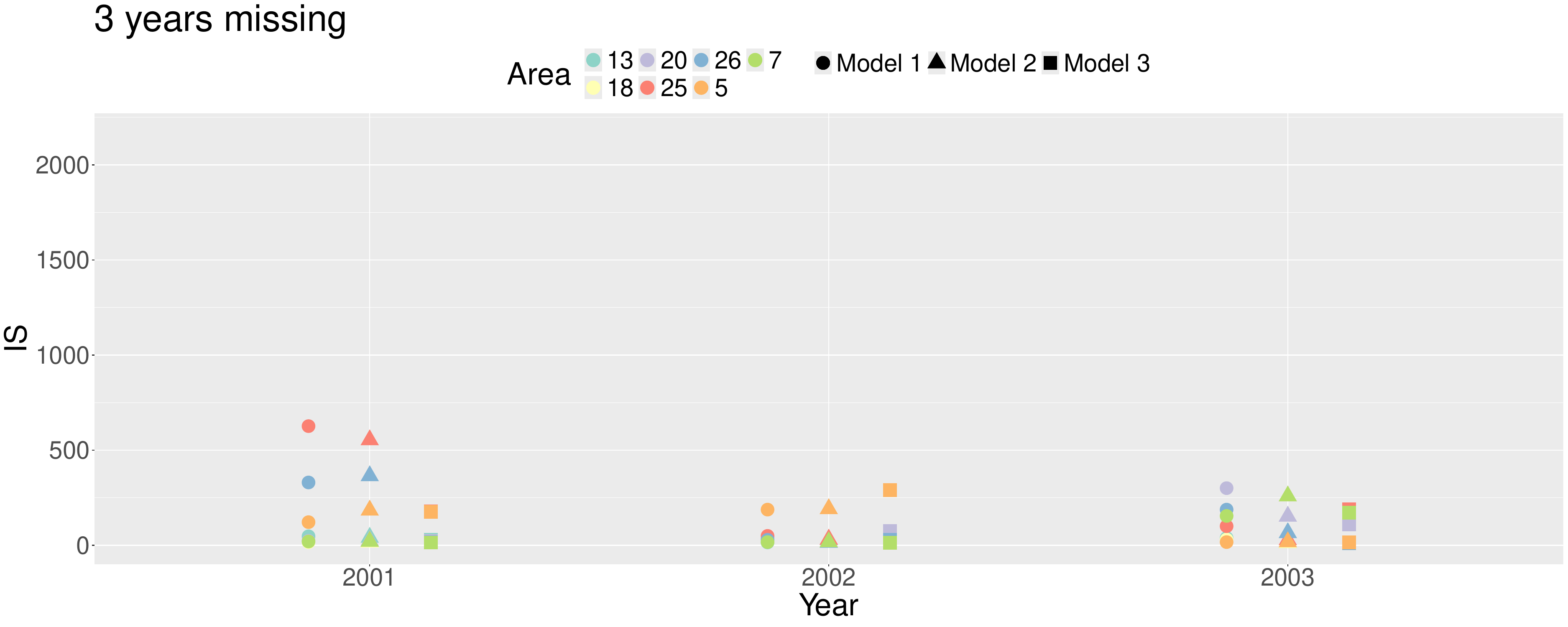}\\ \vspace*{0.5cm}
			
			\includegraphics [width=14cm, page=4]{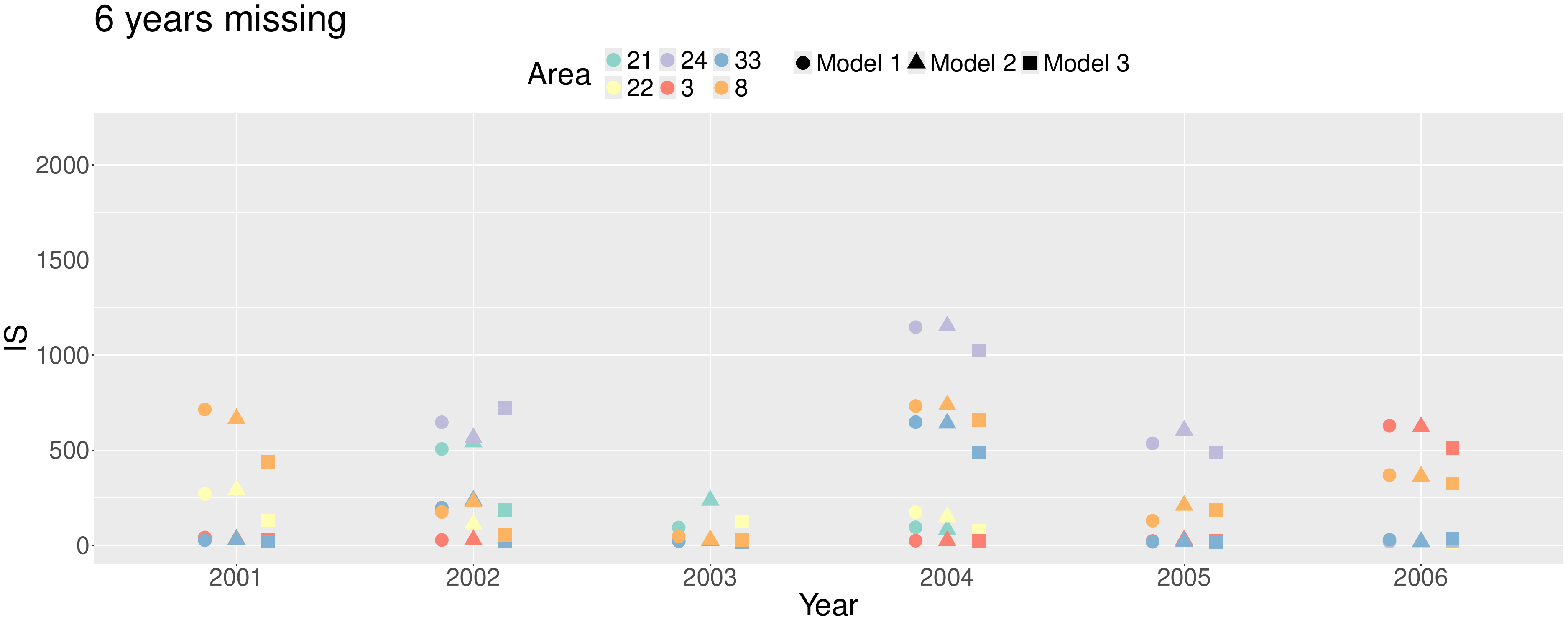}\\ \vspace*{0.5cm}
			
			\includegraphics [width=14cm, page=5]{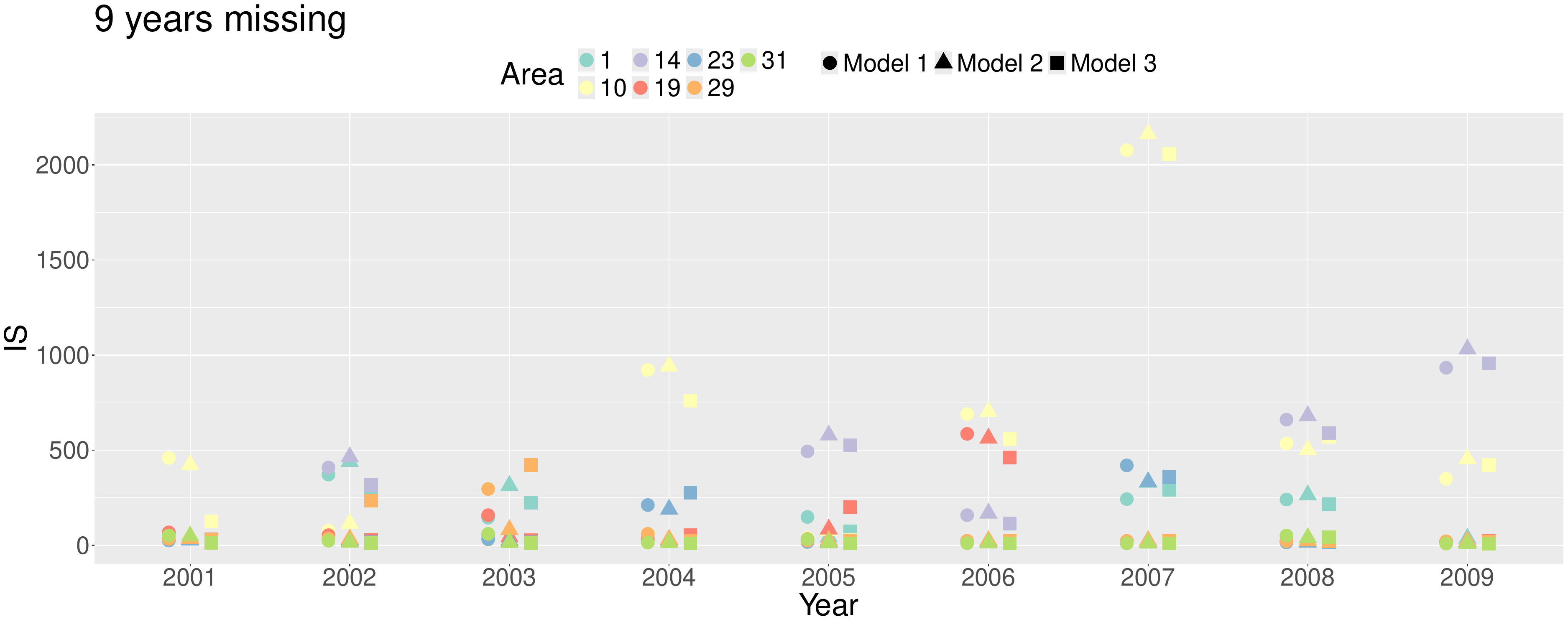}
		\end{center}
		\caption{Average values across validation periods for DSS achieved by each model,  shown for areas with 3 years missing data (top panel), 6 years missing (middle panel), and 9 years missing (bottom panel). Different shapes indicate the models used: dots for Model~1, triangles for Model~2, and squares for Model~3. Colours distinguish different areas within each group. \label{figA3.a} }
	\end{figure}
	
	\begin{figure}[t!]
		\begin{center}
			\includegraphics [width=14cm, page=1]{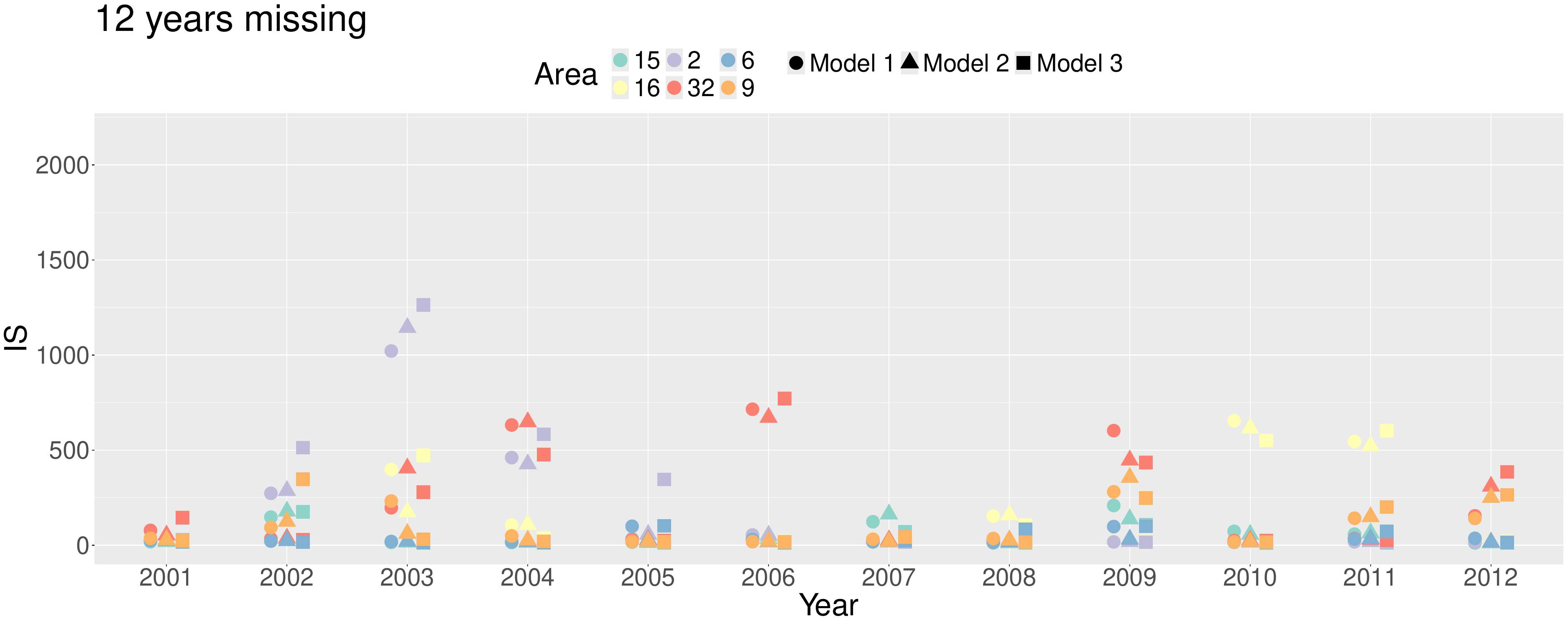}\\ \vspace*{0.5cm}
			
			\includegraphics [width=14cm, page=2]{Fig_SUP/FigureA3.pdf}
		\end{center}
		\caption{Average values across validation periods for DSS achieved by each model,  shown for areas with 12 years missing data (top panel) and 15 years missing (bottom panel). Different shapes indicate the models used: dots for Model~1, triangles for Model~2, and squares for Model~3. Colours distinguish different areas within each group. \label{figA3.b} }
	\end{figure}
	
	\autoref{figA3.a} and \autoref{figA3.b} present the average values across validation periods for IS, achieved by each model, with results shown for areas with 3 years missing data (blue areas in Figure~5 of main paper), 6 years missing (green areas) and 9 years missing (pink areas), and 12 years missing (orange areas) and 15 years missing (garnet areas), respectively. In general, very high IS values are observed, and, similar to the ARB results, no clear pattern emerges among the areas. These results, once again, are influenced by the variability observed in the crude rates.
	
	\clearpage
	\section{Short-term forecasting}\label{AppendixC}
	
	This section presents and discusses figures related to the short-term forecasting that are not included in the main paper. Specifically, it includes the DSS and IS results for one-, two- and three-year-ahead projections, along with the posterior median and 95\% credible intervals for the forecasted rates per 100,000 inhabitants for four selected areas.

	\begin{figure}[b!]
		\begin{center}
			\includegraphics [width=12.5cm]{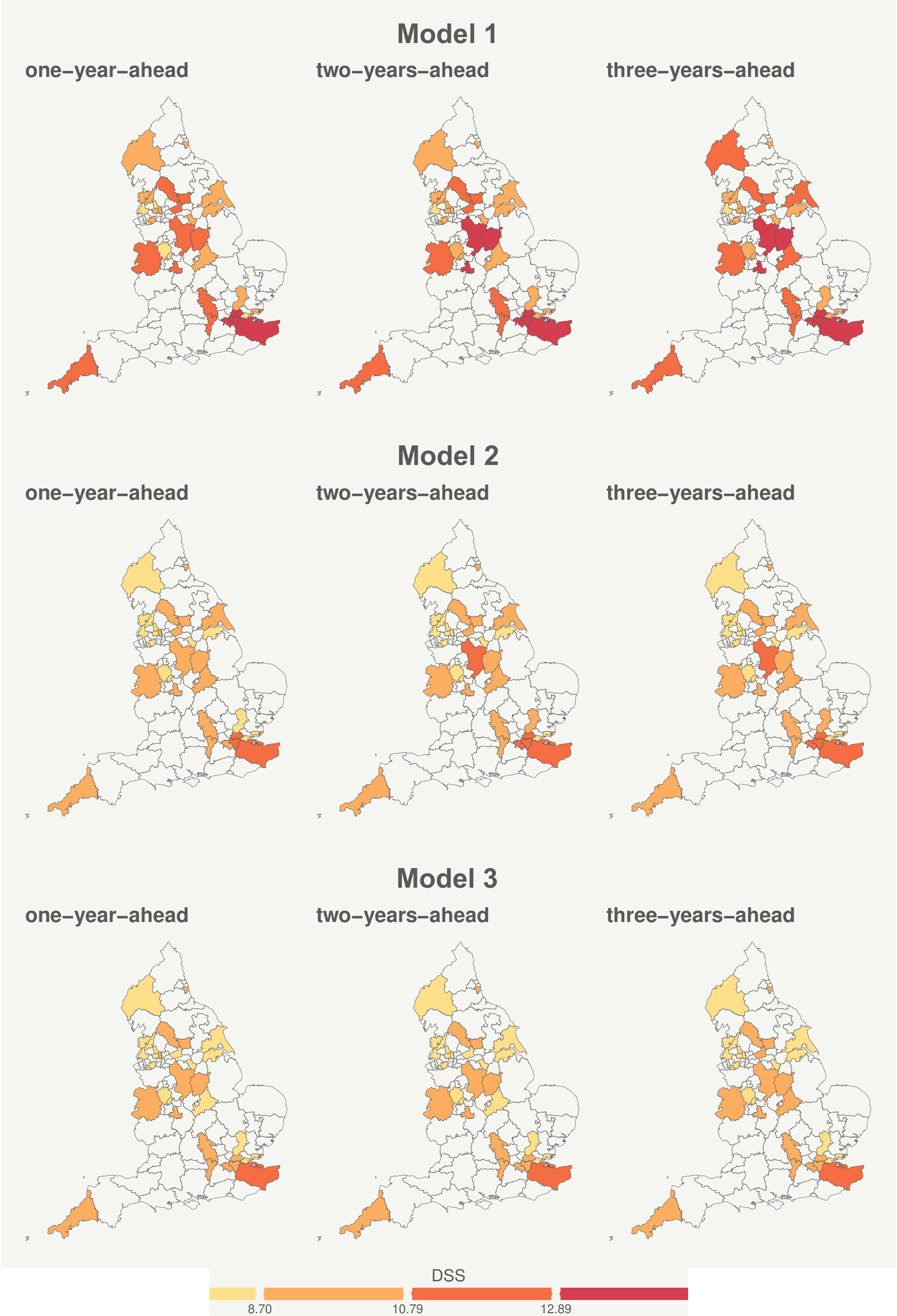}
		\end{center}
		\caption{DSS values for one-, two-, and three-year-ahead predictions are shown for each area with missing data across the time series. The top row displays results using Model~1, the middle row presents results from Model~2, and the bottom row shows the results for Model~3.\label{figA3}}
	\end{figure}
	
	\begin{figure}[t!]
		\begin{center}
			\includegraphics [width=12.5cm]{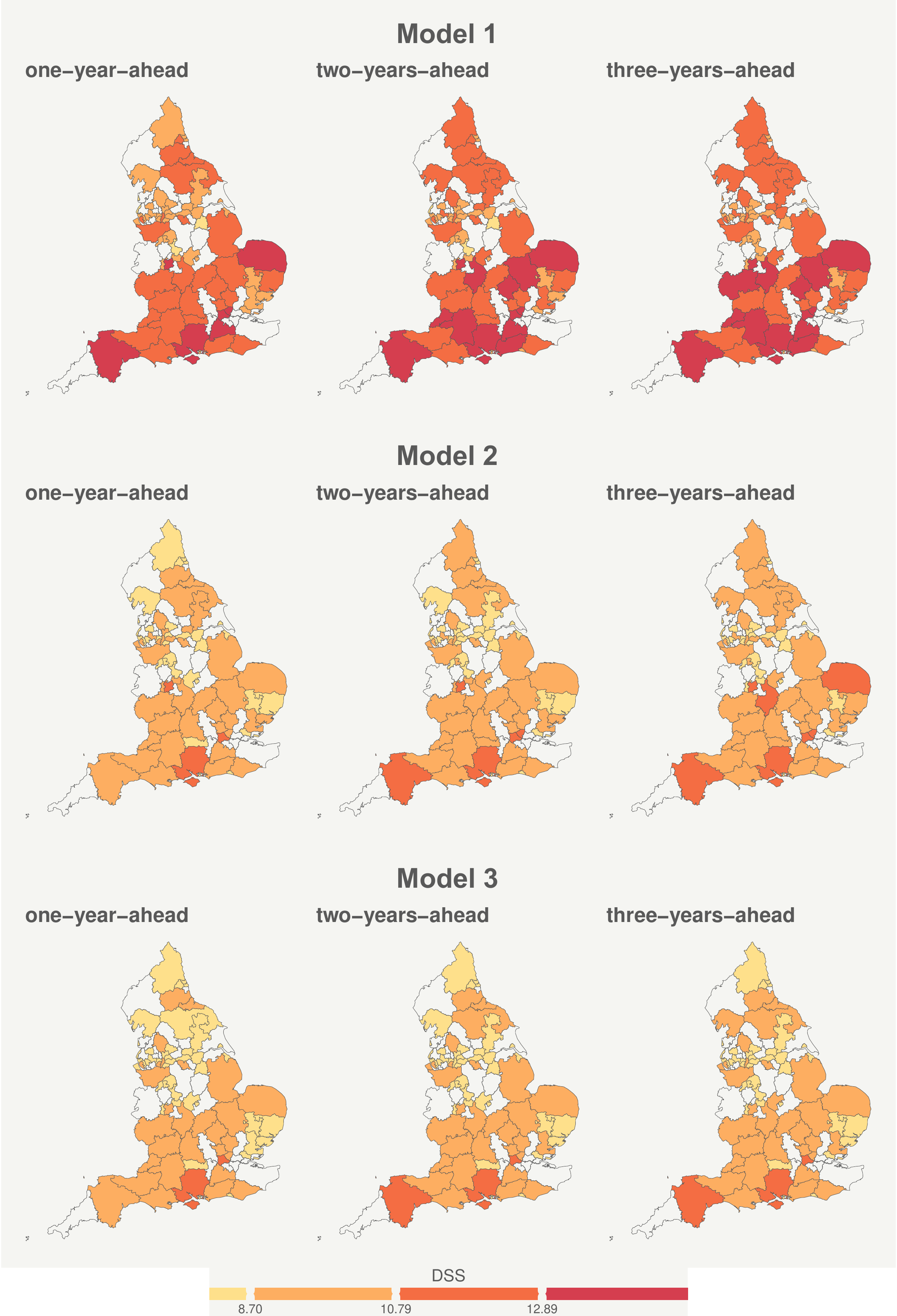}
		\end{center}
		\caption{DSS values for one-, two-, and three-year-ahead predictions are shown for each area with no missing data across the time series. The top row displays results using Model~1, the middle row presents results from Model~2, and the bottom row shows the results for Model~3.\label{figA4}}
	\end{figure}
	
	\autoref{figA3} and \autoref{figA4} depict the DSS values across all regions and forecast horizons (years-ahead predictions) for each of the three proposed models, illustrating areas with missing data and areas with complete time series respectively. The DSS values for both groups, areas with missing data and those with complete series, are similar, suggesting that the number of available years in the series before forecasting has minimal impact on the DSS values.
	The areas with the highest and lowest DSS values obtained by each model remain consistent across all forecast horizons, with a slight increase observed as the forecast horizon extends. Notably, Model~1 consistently achieves the highest DSS values across all areas. While Model~2 and Model~3 produce similar results, Model~3 tends to achieve lower DSS values as the forecast horizon increases, indicating better performance in terms of DSS as the forecast horizon lengthens. \vspace*{0.5cm}

	\begin{figure}[b!]
		\begin{center}
			\includegraphics [width=12.5cm]{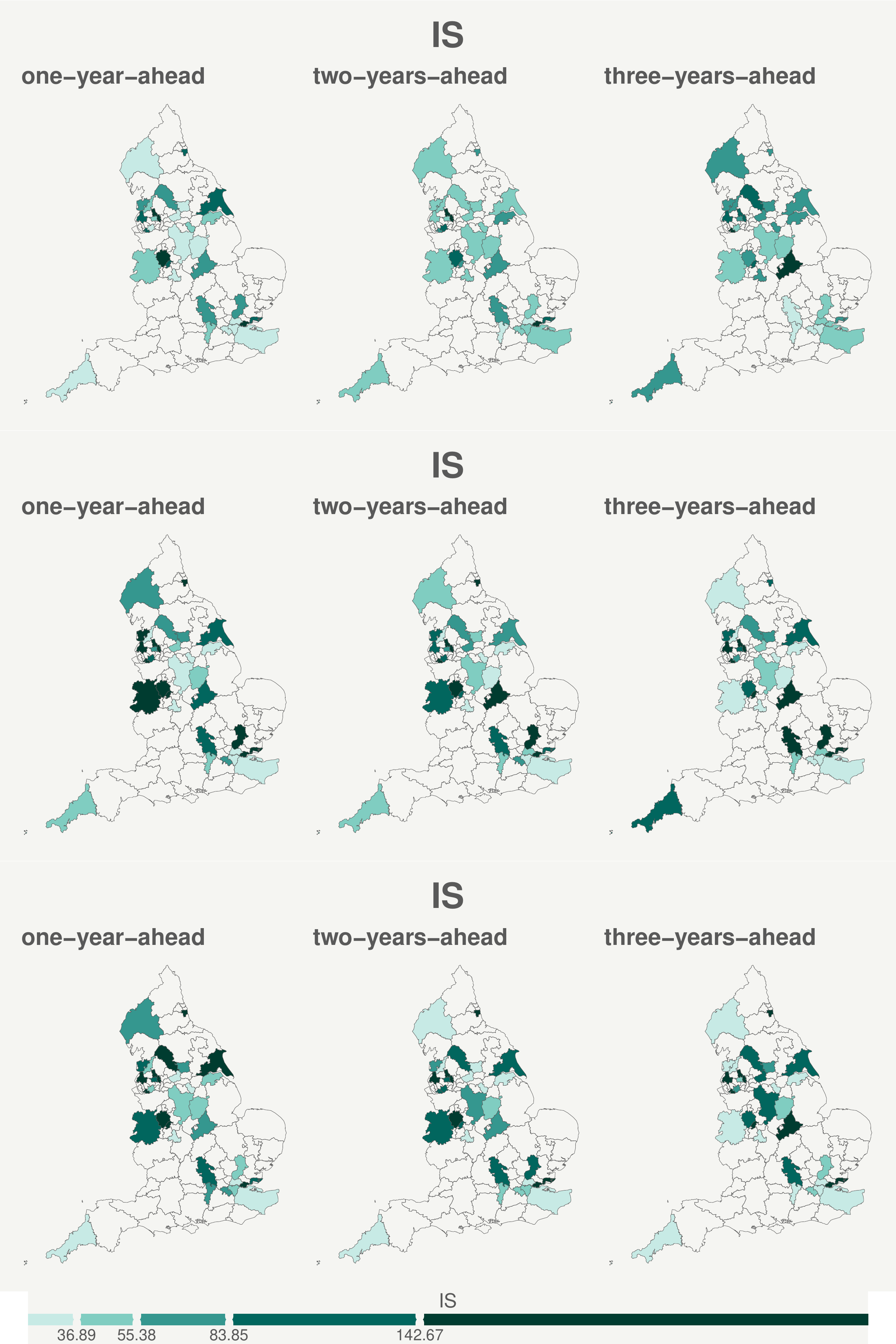}
		\end{center}
		\caption{IS values for one-, two-, and three-year-ahead predictions are shown for each area with missing data across the time series. The top row displays results using Model~1, the middle row presents results from Model~2, and the bottom row shows the results for Model~3.\label{figA5}}
	\end{figure}

	\begin{figure}[t!]
		\begin{center}
			\includegraphics [width=12.5cm]{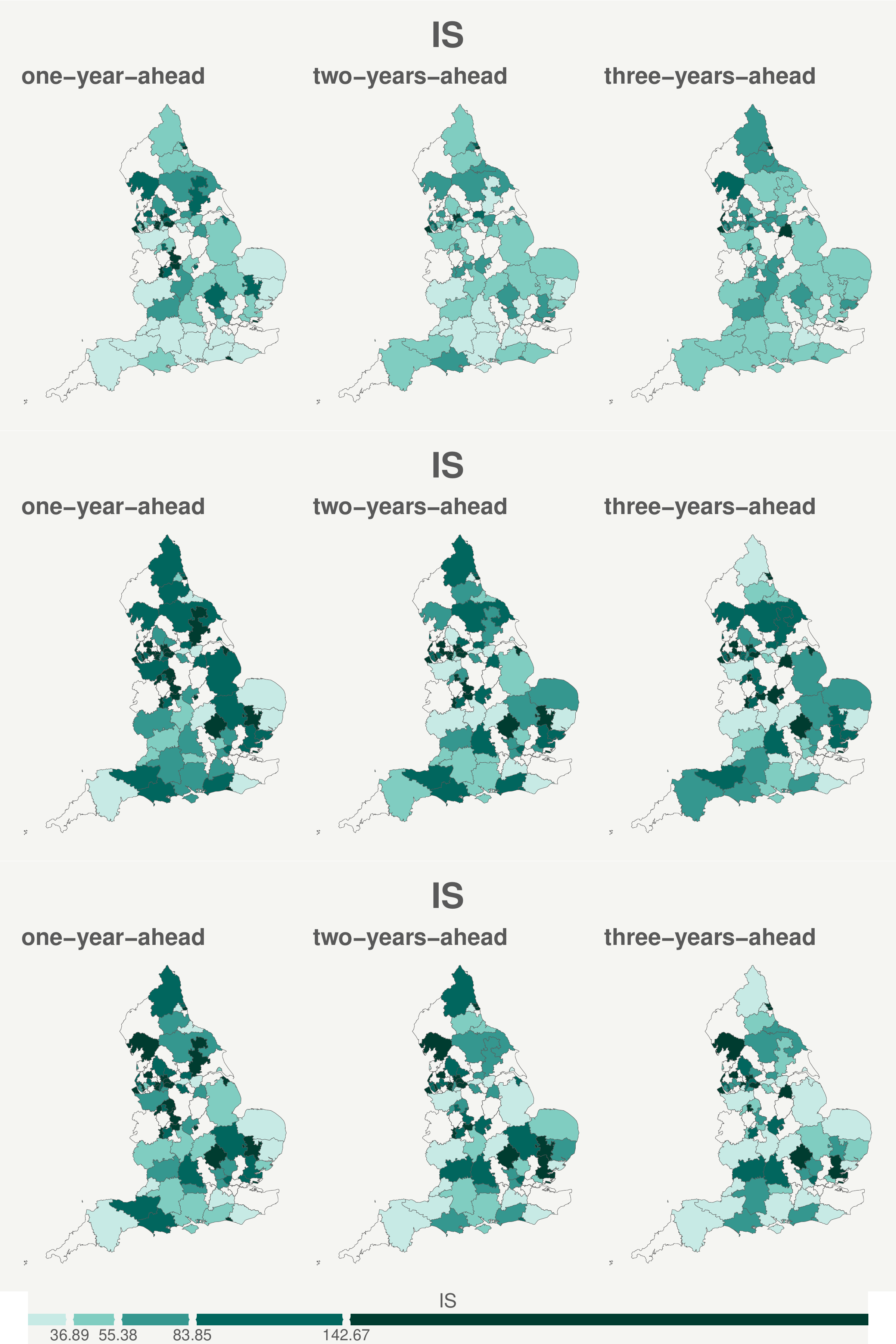}
		\end{center}
		\caption{IS values for one-, two-, and three-year-ahead predictions are shown for each area with no missing data across the time series. The top row displays results using Model~1, the middle row presents results from Model~2, and the bottom row shows the results for Model~3.\label{figA6}}
	\end{figure}
	
	\autoref{figA5} and \autoref{figA6} illustrate the IS values across all regions and forecast horizons for each of the three proposed models, illustrating areas with missing data and areas with complete time series, respectively.
	Areas with missing data and those with complete series exhibit similar IS values, further demonstrating that the number of available years in the series before forecasting has minimal impact on the projections. Notably, there is no significant deterioration in performance as the forecast horizon increases. However, the areas with the highest and lowest IS values obtained by each model vary depending on the forecast horizon, which is reflected in the distinct spatial distribution of IS values across different forecast horizons. Model~2 achieves the highest IS values for most areas, while Model~1 shows more consistent IS values across all areas. In contrast, Model~3 displays greater variability in IS values, with some areas showing very low IS (below 36.89) and others exhibiting very high IS (above 142.67).

	\begin{figure}[b!]
		\begin{center}
			\includegraphics [width=13.5cm]{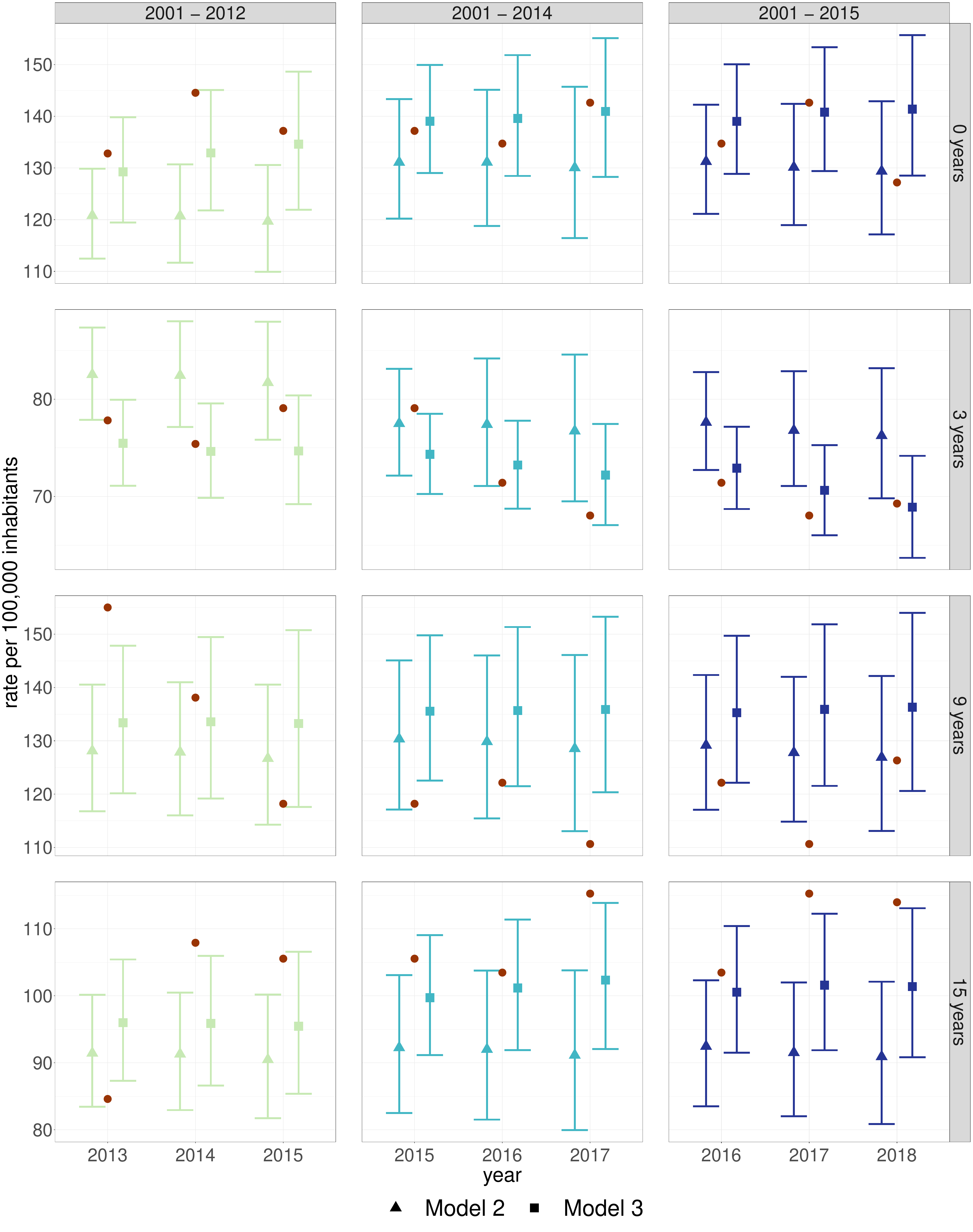}
		\end{center}
		\caption{Posterior median and 95\% credible interval of the forecasted rates per 100,000 inhabitants for four selected areas across validation periods 2001-2012 (left column), 2001-2014 (middle column) and 2001-2015 (right column) obtained by Model~2 (triangles) and Model~3 (squares). Crude rates are represented by brown dots.\label{figA7}}
	\end{figure}
	
	\autoref{figA7}, similar to Figure~11 of the main paper, illustrates the posterior median and the 95\% credible interval of the forecasted rates per 100,000 inhabitants for Model~2 and Model~3 in four selected areas across three validation periods. Due to the large 95\% credible intervals lengths (CILs) of Model~1, it is excluded from this analysis. The areas selected for comparison are the ones depicted in Figure~11 of the main paper; however, in \autoref{figA7} the validation periods shown correspond to  2001-2012, 2001-2014 and 2001-2015. The crude rates are represented by brown dots in \autoref{figA7}. Similar conclusions to those observed in Figure~11 of the main paper are evident. Predictions from Model~2 remain nearly constant across all forecast horizons for each validation period. In contrast, forecasted values from Model~3 show more variability. For example, forecasted rates for 2013-2015 show a clear increase in the area with no missing years, while forecasted rates for 2015-2017 show a decrease in area with three missing years. In general, Model~3 provides more accurate predictions.	
\end{appendices}
	
\end{document}